\documentclass[modern, onecolumn]{aastex62}
\usepackage{graphbox,graphicx}

\usepackage{subfigure}
\usepackage{comment}
\usepackage{threeparttable}
\usepackage{longtable}
\usepackage{afterpage,booktabs}
\usepackage{xcolor}
\usepackage{float}

\received{ddmmyyyy}
\revised{ddmmyyyy}
\accepted{ddmmyyyy}
\submitjournal{ApJ}

\shorttitle{Classification of Photospheric Emission in Short GRBs}
\shortauthors{Dereli-B\'egu\'e et al.}

\begin{document}

\title{Classification of Photospheric Emission in Short GRBs}

\correspondingauthor{H\"usne Dereli-B\'egu\'e}
\email{husnedereli@gmail.com}


\author[0000-0002-8852-7530]{H\"usne Dereli-B\'egu\'e}
\affil{Max Planck Institute for Extraterrestrial Physics, Giessenbachstrasse 1, D-85748 Garching, Germany}
\affil{Department of Physics, Bar-Ilan University, Ramat-Gan 52900, Israel}

\author[0000-0001-8667-0889]{Asaf Pe'er}
\affil{Department of Physics, Bar-Ilan University, Ramat-Gan 52900, Israel}

\author[0000-0002-9769-8016]{Felix Ryde}
\affil{Department of Physics, KTH Royal Institute of Technology, \\
and The Oskar Klein Centre, SE-106 91 Stockholm, Sweden 
}


\begin{abstract}

In order to better understand the physical origin of short duration gamma-ray bursts (GRBs), we perform time-resolved spectral analysis on a sample of 70 pulses in 68 short GRBs with burst duration $T_{90}\lesssim2$~s detected by the \textit{Fermi}/GBM.
We apply a Bayesian analysis to all spectra that have statistical significance $S\ge15$ within each pulse and apply a cut-off power law (CPL) model. We then select in each pulse the timebin that has the maximal value of the low energy spectral index, 
for further analysis. Under the assumption that the main emission mechanism is the same throughout each pulse, such an analysis is indicative of pulse emission. We find that $\sim$~1/3 of short GRBs are consistent with a pure, non-dissipative photospheric model, at least, around the peak of the pulse. This fraction is larger compare to the corresponding one (1/4) obtained for long GRBs.
For these bursts, we find (i) a bi-modal distribution in the values of the Lorentz factors and the hardness ratios; (ii) an anti-correlation between $T_{90}$ and the peak energy, $E_{\rm pk}$: $T_{90} \propto E_{\rm pk}^{-0.50\pm0.19}$. This correlation disappears when we consider the entire sample. Our results thus imply that the short GRB population may in fact be composed of two separate populations: one being a continuation of the long GRB population to shorter durations, and the other one being distinctly separate with different physical properties. Furthermore, thermal emission is initially ubiquitous, but is accompanied at longer times by additional radiation (likely synchrotron).

\end{abstract}



\section{Introduction}
\label{sect:intro}
After more than four decades of extensive research, the physical origin of gamma-ray burst (GRB) prompt spectra remains unclear and highly debated. The classification of GRBs is a tool which could help to understand the emission mechanisms at work. The main classification into short-hard and long-soft is based on their duration and spectral hardness. The short GRBs have a duration shorter than 2~seconds while the long GRBs have a duration longer than 2~seconds \citep{Kouveliotou1993}. The spectral peak energy of short bursts is, on the average, higher than that of long GRBs \citep[e.g.,][]{Ghirlanda2011}.
However, both classes share many spectral characteristics, for instance, their spectra peak in the MeV range, with power law extensions below and above the peak.
Both populations have a common inverse correlation between the intensity and the duration for individual pulses \citep{Hakkila2011, Norris2011}, and they follow a similar relation between the peak energy, $E_{\rm pk}$ and the peak luminosity, $L_{\rm peak}$ as well as the isotropic equivalent energy, $E_{\rm iso}$ \citep{Yonetoku2004, Amati2006, Ghirlanda2009}.

Despite these observed similarities, short and long bursts are thought to originate from different progenitors; the collapse of a very massive star for long GRBs \citep{Woosley1993} and a compact binary merger for short GRBs \citep{Eichler1989}. In fact, long GRBs are studied more than short ones. Indeed, they release more photons which allows more detailed spectral studies. In addition, more redshifts are known for long GRBs than for short ones since the afterglow after a few thousands seconds is brighter for long bursts. This allows the study of intrinsic properties \citep[e.g.,][]{Howell2013}. The recent increase of interest in the study of short GRBs is mostly due to the detection of short GRB 170817B simultaneously with the first gravitational wave (GW) from a merger of binary neutron stars \citep{Abbott2017, Goldstein2017}.  

Observationally, many GRB prompt spectra have too narrow $\nu F_{\nu}$ peaks compared to what is expected from the synchrotron emission model \citep[e.g.][]{Ryde2004, Axelsson2015, Yu2015a}. Yet, they are broader than a Planck spectrum \citep{Goodman1986, Paczynski1986, Beloborodov2011}. Photospheric emission from highly relativistic outflows is often used to explain this observed spectral shape. Broadening of the spectrum by energy dissipation below the photosphere can be caused by shocks, dissipation of magnetic energy or collisional processes \citep{Giannios2005, Peer2006, Beloborodov2010}. Moreover, broadening in a passively cooled jet without any energy dissipation can be due to geometrical broadening occurring during the coasting phase \citep{Beloborodov2011, Begue2013, Lundman2013}. In order for the emission to be detectable the outflow has to become transparent below or close to the saturation radius, $r_s$, where the outflow saturates to its final outflow Lorentz factor \citep{Meszaros2006, Ryde2017}.

The observed spectral shape of the prompt emission is commonly characterised by empirical models, such as the "Band" model \citep{Band1993} or a cutoff power-law model \citep[see, e.g.,][]{Yu2016}. However, in making the link between observation and theory, the parameters of the empirical models should not be used directly for the comparison with the prediction of physical emission models. Indeed, attempt to make such a link leads to two main problems. The first one is known as an energy-window bias effect. When the empirical model does not match the true spectral shape (its curvature where a spectral peak lies inside the GBM energy window) then physical interpretation of the model parameters will be wrong; e.g, there will be a positive correlation between the parameters of the empirical model at low peak energies \cite[e.g., ][]{Preece1998, Lloyd2000, Burgess2015, Ryde2019, Acuner2019}. The second problem is the limitation due to the band-width of the detector which prevents the full spectrum to be detected \citep{Burgess2015, Ryde2019}. 

There are two solutions to overcome these problems. The first one is to use a physically motivated model and fit it directly to the data \citep[e.g.,][]{Lloyd2000, Ahlgren2015}. In this way, there is no need for an empirical function. However, it is computationally expensive due to the need to make a forward-folding of the theoretically generated spectra through the detector's response matrix, and the need to subtract the background - both vary from burst to burst. Thus, the claimed model has to be fitted individually to each burst. Furthermore, one has to assume knowledge of the physical model that should be used \citep[e.g.,][]{Baring2004, Burgess2016, Burgess2019b, Oganesyan2018}. Due to these limitations, this direct method was applied, so far, only to a limited number of bursts \citep[e.g.,][]{Vianello2018b, Burgess2011, Ahlgren2019}.

The second solution is to use an assumed physical model to generate a large number of synthetic spectra which are, in turn, fitted with empirical functions. This provides the distribution of the empirical model parameters that the given theoretical model corresponds to. The properties of the parameter distributions, for instance their widths, depend on how well the empirical model matches the theoretical model. These distributions can then be compared to the full GRB catalog, in order to assess the theoretical model's ability to explain the data. This method was used by several authors \citep[e.g.,][]{Burgess2015, Acuner2019} to make statistical claims about the ability of a theoretical model to fit the data.     

In an attempt to fit a non-dissipative photospheric model \citep{Beloborodov2011, Lundman2013} to GRB spectra, \citet{Acuner2019} followed the second method and generated a series of synthetic spectra with a high signal-to-noise ratio (SNR) of 300 and peak energies at the range of 40-2000 keV. 
The simulated (synthetic) spectral data were fitted with a cutoff power-law model. It was found that the distribution of the low-energy photon indexes ranges from -0.4 to 0.0 and peaks at around -0.1. This was then compared with the distribution of the maximal, time-resolved value of the low energy spectral index, $\alpha_{\rm max}$, in the samples of \citet{Yu2016, Yu2019}. They found that 1/4 of the long bursts have an $\alpha_{\rm max}$ which is consistent with a non-dissipative outflow, releasing its thermal energy at the photosphere. However, \citet{Acuner2019} did not consider short bursts since the selection criteria of \citet{Yu2016, Yu2019} is mainly based on bright burst with duration $T_{90}$ $\gtrsim 2$ s.

While the spectral properties of short GRBs are much less studied than that of long GRBs, evidence are accumulating that photospheric (thermal) emission could play an important role in these bursts as well. 
The main motivation for our current study is the large number of short GRBs seen in the cluster 5 in \citet{Acuner2018}. This cluster was found to be consistent with a photospheric emission origin. Therefore, in this work, we are also using the fitted synthetic spectra from \citet{Acuner2019} to find the fraction of short GRBs compatible with a non-dissipative photosphere (NDP) model. As a first step, we apply time-resolved analysis to the spectra of individual pulses obtained from 68 short GRBs and use a Bayesian analysis approach. As a second step we study, in detail, the timebins with the hardest low-energy spectral index ($\alpha_{\rm max}$) in each pulse.

This paper is organised as follows. In Section \ref{sect:sample_selection}, we define the sample of short GRBs and present the analysis methods. In Section \ref{sect:spectral_results}, we present result of the spectral parameter relations, the observed $\alpha$ distributions, the Lorentz factor for the bursts consistent with thermal emission, and the hardness ratio. In Section \ref{sec:Discussions}, we then discuss our choice of spectral fitting model, and the correlations between temporal and spectral structures. 
Finally, in Section \ref{sec:Summary_Conclusion} we list our summary and conclusions.


\section{Data Collection and Analysis Method}
\label{sect:sample_selection}

\subsection{Sample Selection}
\label{sect:sample}
We select short GRBs, namely GRBs having duration $T_{90} < 2$ seconds detected by the Gamma-ray Burst Monitor (GBM) onboard the \textit{Fermi Gamma-ray Space Telescope} during the first 11 years of its mission. We scan all the short bursts for which automatic spectral fits are performed on the time-resolved data around the peak flux, within the time interval given in the GBM catalog \citep{VonKienlin2014}. We find a total of 147 short bursts for which spectral fits can be carried out and analysed. All the data is taken from the \textit{Fermi}/GBM burst catalog published at HEASARC~\footnote{\url{https://heasarc.gsfc.nasa.gov/W3Browse/fermi/fermigbrst.html}}. We further set a limit for at least one timebin to have a statistical significance (see Section \ref{sec:method} for the definition), $S\ge15$ in each pulse; we end up having 70 pulses from 68 short GRBs as a final sample listed in Table \ref{tab:sample1}.


\subsection{Analysis Method}
\label{sec:method}
For the analysis, we follow the procedure of the \textit{Fermi}/GBM GRB time-integrated \citep{Goldstein2012,Gruber2014} and time-resolved catalogs \citep{Yu2016,Yu2019}. We select at most three NaIs and one BGO for the spectral analysis of each short GRB, see in Table \ref{tab:sample1}, Column 3. We use the response files RSP (except for two cases, GRB 090510, GRB 170127, in which the RSP2 files are used) for each short GRB.   
We further use the standard \textit{Fermi}/GBM energy ranges: $8-30$~keV and 40~keV to $\sim$850~keV for the NaI detectors (avoiding the K-edge at 33.17~keV)\footnote{\url{https://fermi.gsfc.nasa.gov/ssc/data/analysis/GBM_caveats.html}}, and $\sim$250~keV to 40~MeV for the BGO detectors. 

We select the source interval from the first few seconds of the burst light curve where the first pulse is most prominent, see in Table \ref{tab:sample1}, Column 4. Indeed, most of the bursts in the sample are single pulsed bursts. We use the NaI detector in which the largest photon counts per second was recorded from the burst to define the background intervals before and after the pulse, see in Table \ref{tab:sample1}, Columns 3, 5 and 6 respectively. These intervals are then applied to all detectors. As a standard procedure in GRB background fitting of GBM data, we fit a polynomial with the order of between 0 and 4, to the total count rate of each energy channel (128 channels for TTE) of each detectors. From this fit the optimal order of the polynomial is determined by a likelihood ratio test. Then, this order of polynomial is interpolated through the source time interval to estimate the background photon count flux and its corresponding errors in each energy channel during the time of source activity.

We then rebin the light curves by applying the Bayesian block method \citep{Scargle2013} to the unbinned TTE data. This method  identifies intervals that are consistent with a constant Poisson rate. The light-curves are thus rebinned into intervals over which the intensity change is small. 
The method uses a probability of a false positive of an intensity change $p_0$, which we set to $p_0 = 0.01$ \footnote{
 The exact value of $p_0$ is determined through a tradeoff between the risk of identifying noise fluctuations versus missing real intensity changes. It needs to be found through an iterative method, but is not very sensitive for data with even moderate significance \citep{Scargle2013}.
 }.
This is the typical value employed for GBM data analysis \citep[e.g.][]{Vianello2018b, Burgess2019a, Yu2019}. 
A consequence of the Bayesian block method is that the timebins will have variable widths and variable statistical significance. However, it ensures that the emission evolution is small within a timebin, which is essential in order to capture the instantaneous emission spectrum. We use the TTE data of the brightest NaI detector and its binning is then transferred and applied to all other detectors. The total number of bins for each pulse is listed in Table \ref{tab:sample1}, Column~7. 

We further estimate the  significance of the signal in  each timebin. We employ the significance $S$ given in equation (15) in \citet{Vianello2018a}, which is suitable for Poisson sources with Gaussian backgrounds. In particular, it is applicable for our analysis of GBM data, because the background is not measured in an off-source interval, but is estimated through a polynomial fit as described above.  In such a case, the typically employed signal-to-noise-ratio (SNR), given by $(n-b)/\sqrt{b}$ where $n$ is the measurement and $b$ is the background estimate, is not strictly valid and typically overestimates the significance measure \citep{Vianello2018a}. We find that the parameters from spectral modelling are typically well constrained when the statistical significance $S\ge15$ (see appendix \ref{app:significance_level} and \citet{Yu2019} for further details). Therefore, we limit our analysis to timebins with at least this level of significance.

For the spectral model, we use the cutoff power law model (CPL), which is a power law with an exponential cut-off, and that has been extensively used since it is the best model for most GRBs in the Fermi GBM catalogs \citep{Goldstein2012,Gruber2014,Yu2016,Yu2019}. 
The CPL fit parameters are the normalization $K \,(\rm ph~\rm s^{-1}~\rm cm^{-2}~\rm keV^{-1}$), the low-energy power-law index $\alpha$, the cut-off energy $E_{\rm c} \,(\rm keV)$. We further derive the CPL peak energy $E_{\rm pk}$ (keV) and the CPL energy flux $F \, (\rm erg~\rm s^{-1}\rm cm^{-2})$.

To perform the time-resolved spectroscopy, we use the Multi-Mission Maximum Likelihood 3ML package \citep{Vianello2015} and follow the method outlined in \citet{Yu2019}. The spectral analysis is performed with a Bayesian approach. For the likelihood function we use a Poisson distribution for the source signal and a Gaussian distribution for the background signal. We employ prior distributions for the parameters which are based on the parameter ranges found in the GBM catalogs (e.g. \citet{Yu2016}). While the normalization ($K\sim 10^{-11}$---$10^3 ~\rm ph~\rm s^{-1}~\rm cm^{-2}~\rm keV^{-1}$) is assumed to have a log uniform prior, the low energy index ($\alpha \sim -3$ --- 2) and the cut-off energy ($E_{\rm c} \sim 10$ --- $10000 \rm ~keV$) are assumed to have uniform priors. 
We also investigate the sensitivity to the prior choices, by trying different distribution of the priors. We find that the fit results are insensitive to the choice of prior distributions, mainly due to the high significance level of the data that we are analysing. A similar conclusion was drawn by \citet{Acuner2020}.

The spectral analysis yields posterior probability distributions of the parameters by using their prior probability distributions and the likelihood function obtained from the data. For this we used the technique of Markov Chain Monte Carlo (MCMC).
All parameter uncertainties quoted in this paper are characterized by the highest posterior density credible intervals.


\section{Spectral analysis Results}
\label{sect:spectral_results}

\subsection{Parameter Relations: the entire sample}
\label{subsec:global_relations}
\begin{figure*}
\centering
\subfigure{\includegraphics[align=t,width=0.45\linewidth]{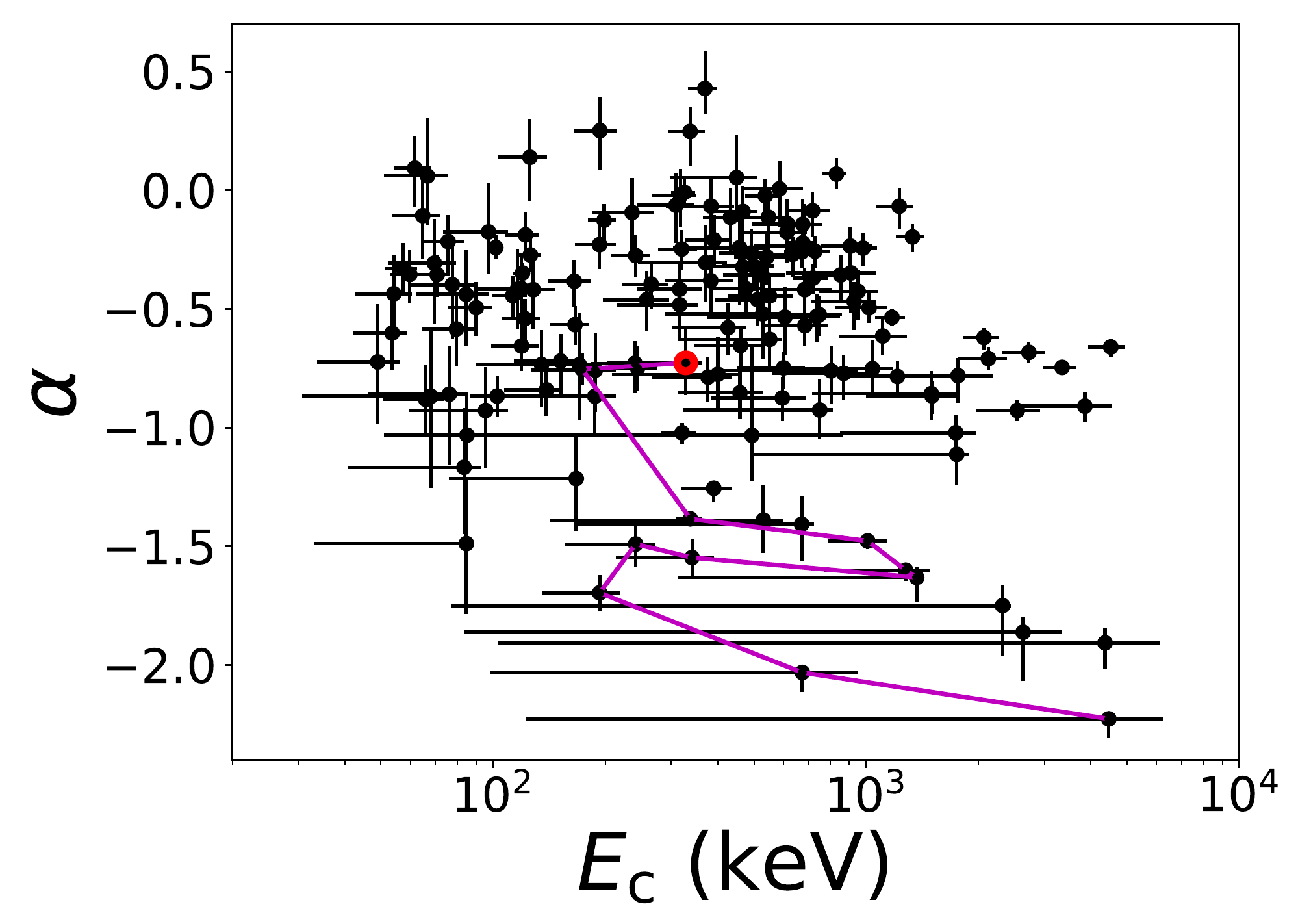}}
\subfigure{\includegraphics[align=t,width=0.45\linewidth]{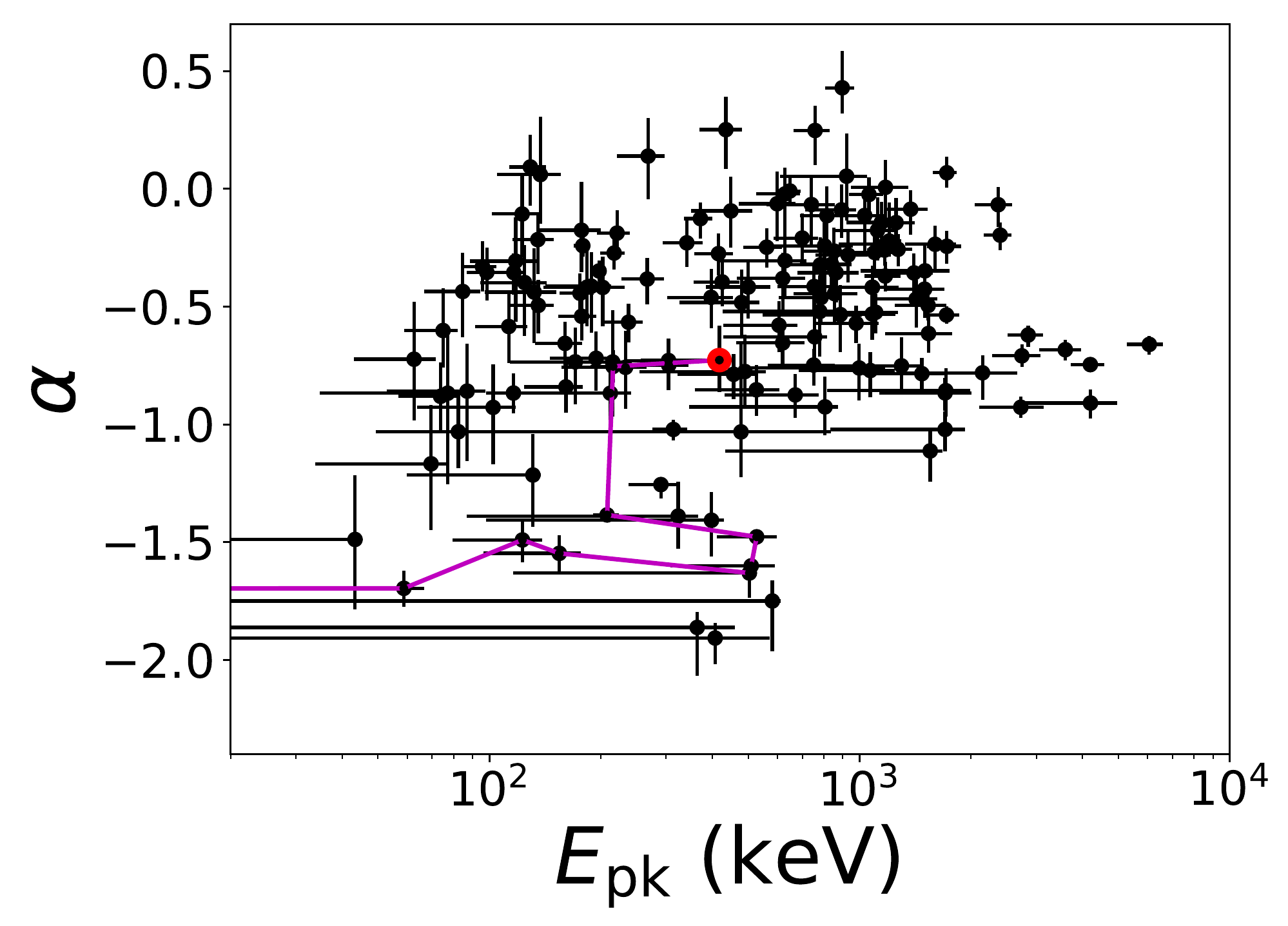}}
\subfigure{\includegraphics[align=t,width=0.45\linewidth]{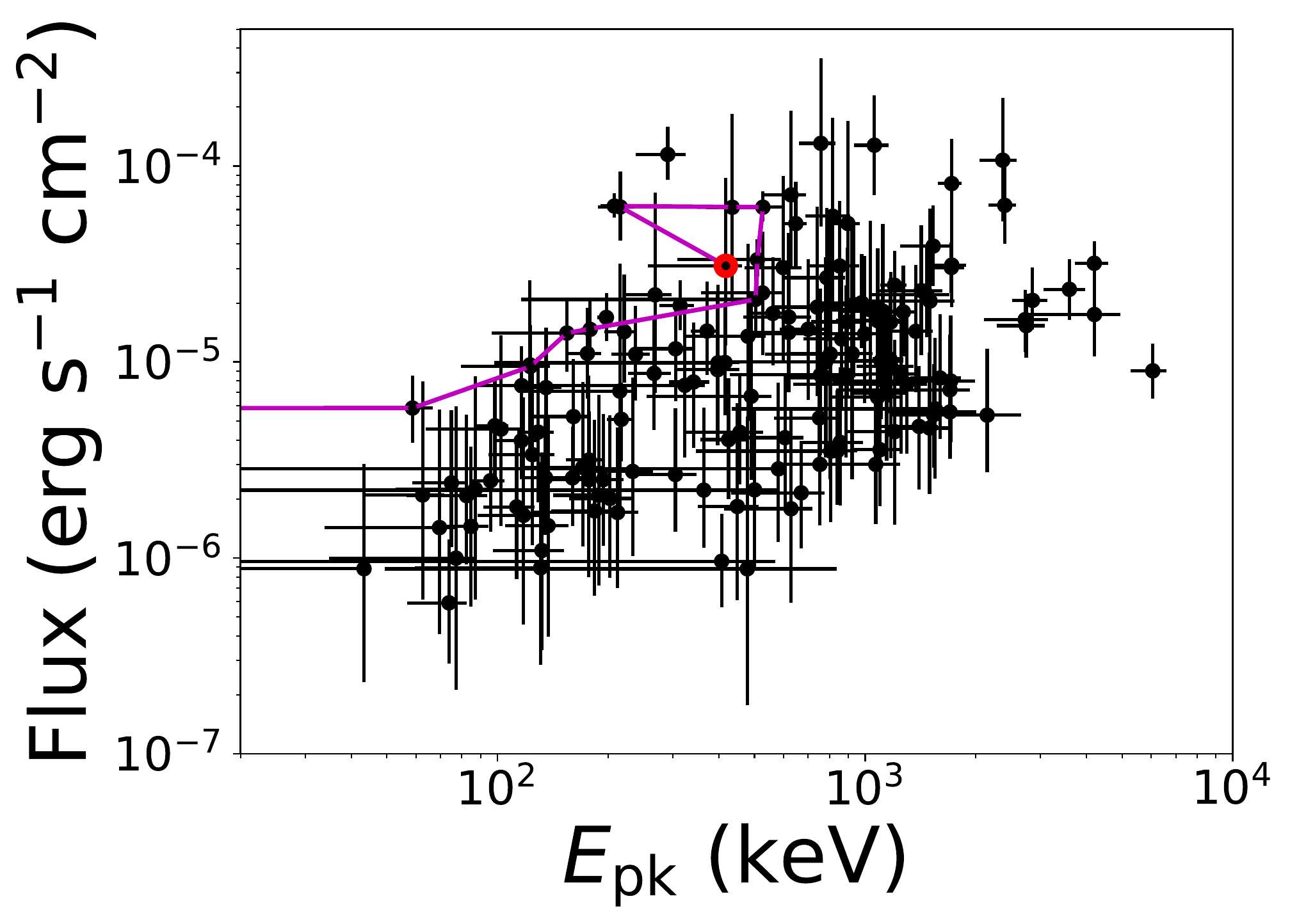}}
\subfigure{\includegraphics[align=t,width=0.45\linewidth]{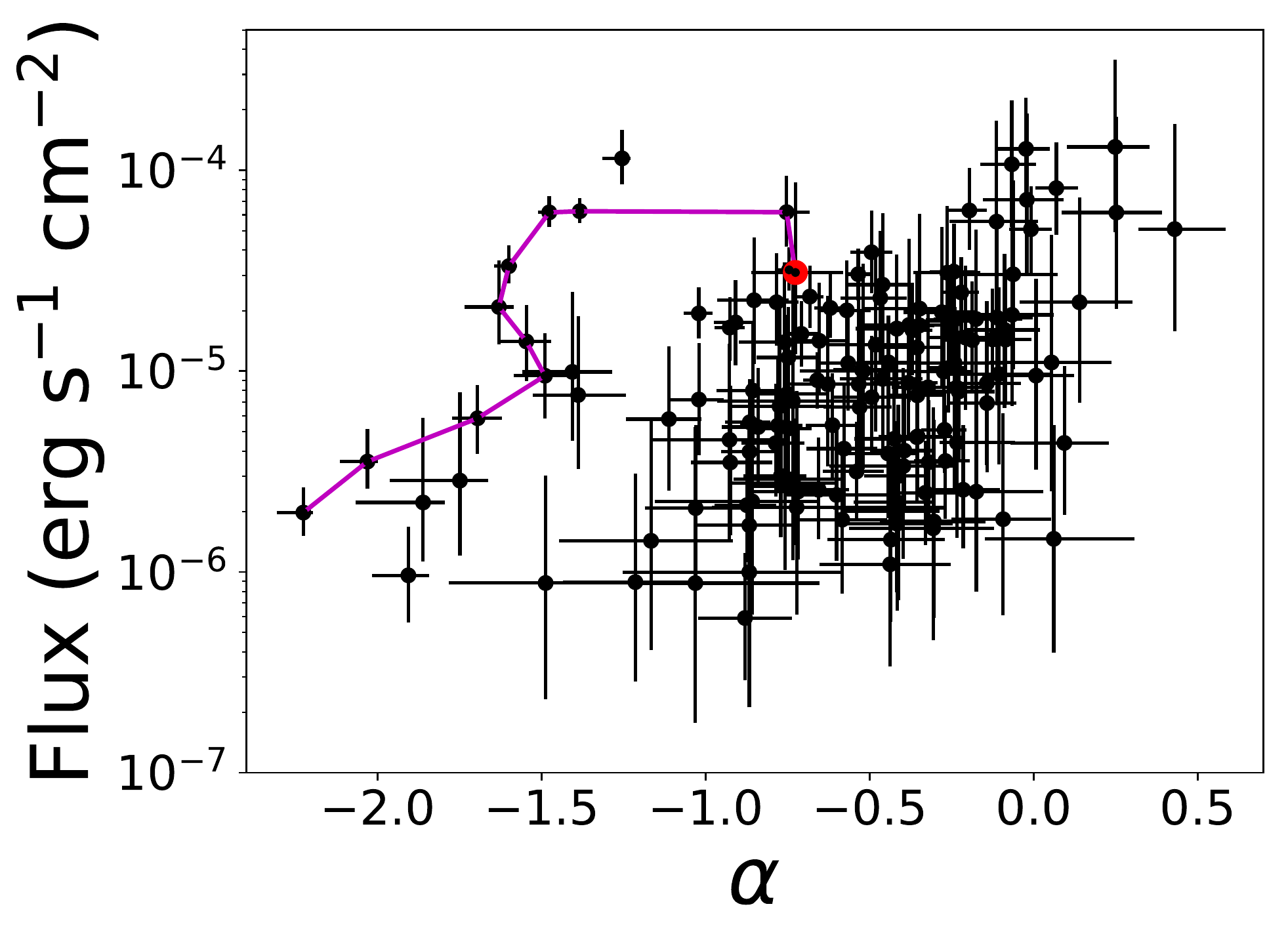}}
\caption{Global relations of the fitted parameters within the GBM energy range (8~keV-40~MeV) for timebins with statistical significance of $S\geq15$. For the definition of parameters see subsection \ref{sec:method}. We only see a trend of a positive correlation between $F$ - $E_{\rm pk}$. The purple lines follow the time evolution of the brightest short GRB 120323 starting with first timebin in red.
\label{fig:correlations_ALL}}
\end{figure*}

Of the 68 short GRBs listed in Table \ref{tab:sample1}, we identified 70 distinct pulses. When dividing into separate timebins, there are total of 475 spectra within the GBM energy range (8 keV to 40 MeV)
that can be analyzed. Out of those, 153 spectra have statistical significance $S\geq15$. The results of the CPL model fits to those 153 spectra are presented in Figure \ref{fig:correlations_ALL}. We show several of the relations: $\alpha-E_{\rm c}$ (upper left panel), $\alpha-E_{\rm pk}$ (upper right panel), $F-E_{\rm pk}$ (bottom left panel), $F-\alpha$ (bottom right panel). 
We only observe a trend of a positive correlation between $F-E_{\rm pk}$ (Figure \ref{fig:correlations_ALL}, bottom left panel)~\footnote{If we remove the brightest short GRB 120323 we do see a weak correlation between $F$ - $\alpha$, with $\alpha$ the Spearman's rank correlation coefficient $r = 0.4$, i.e. the chance probability is $p = 8.4\times10^{-7}$.}.
 
We point out that most of the data with a very low spectral index $-2.1 < \alpha < -1.1$  belong to a single burst, the brightest short GRB 120323 (see Figure \ref{fig:correlations_ALL}). As a result the values of $\alpha$ for the vast majority of bursts in our sample lie between -1.6 and +0.6 within 1-$\sigma$ uncertainty. This range is narrower than that observed in long GRBs, $-2<\alpha<1$ \citep{Yu2019}. However, the range of $E_{\rm pk}$ ($40{\rm~keV}<E_{\rm pk}<6{\rm~MeV}$) and flux ($8\times10^{-6}{\rm~erg~s}^{-1}{\rm~cm}^{-2}<F<9\times10^{-3}{\rm~erg~s}^{-1}{\rm~cm}^{-2}$) obtained from short GRBs are similar to the range obtained in long GRBs, $10{\rm~keV}<E_{\rm pk}<7\times10^3{\rm~keV}$ and $10^{-7}{\rm~erg~s}^{-1}{\rm~cm}^{-2}< F<5\times10^{-5}{\rm~erg~s}^{-1}{\rm~cm}^{-2}$ \citep{Yu2019}.

When comparing the results presented in Figure \ref{fig:correlations_ALL} to the results obtained by \citet{Yu2019}, we conclude that most of the bins obtained from short GRBs generally have harder values of the spectral index $\alpha$, higher energies and have higher fluxes than those in long GRBs.

\subsection{Distribution of low energy spectral indexes}
\label{subsect:distribution}

\begin{figure*}
\centering
\subfigure{\includegraphics[align=t,width=0.45\linewidth]{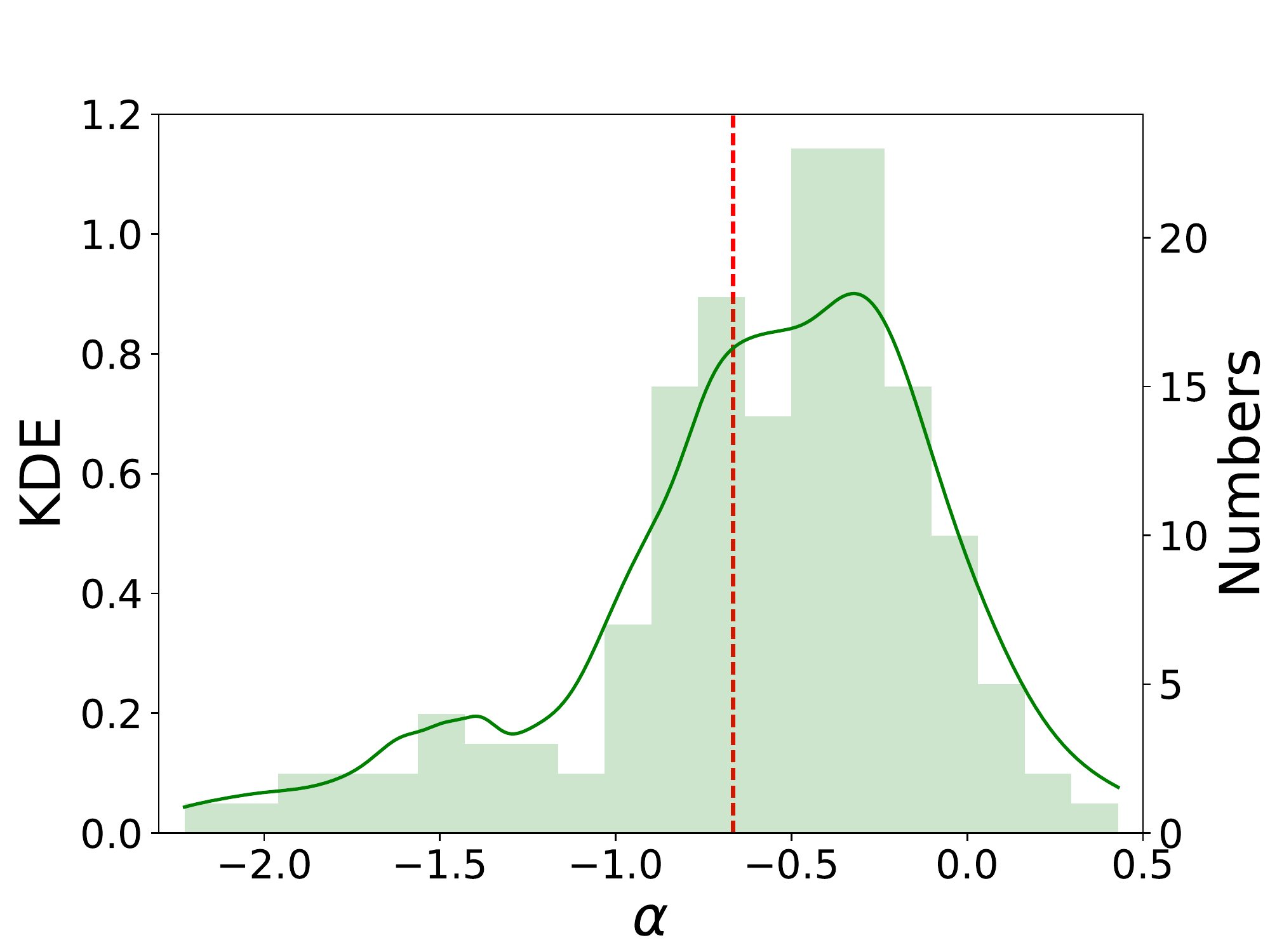}}
\subfigure{\includegraphics[align=t,width=0.45\linewidth]{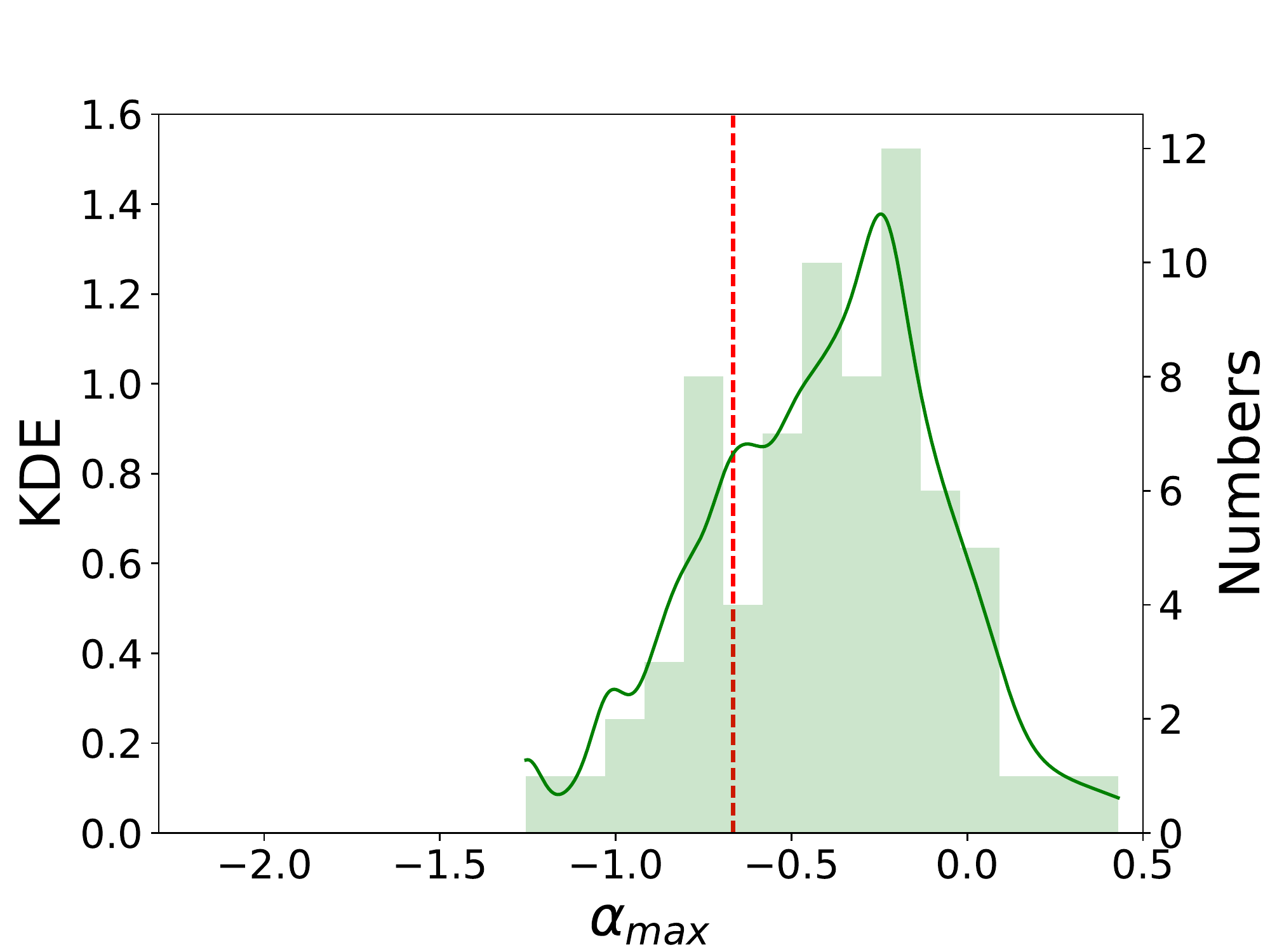}}
\caption{Distribution of the spectral index $\alpha$ of two different samples. Left panel: The $\alpha$-distribution from 153 spectra obtained from the 70 pulses in 68 short GRBs. Right panel: Distribution of the the maximal (hardest) value of $\alpha$, denoted $\alpha_{\rm max}$, in each of the 70 pulses. The red dashed lines indicate the line-of-death, $\alpha = -2/3$, for synchrotron emission. In both panels the right-hand ordinate is the number of spectra in each histogram bin and the left-hand ordinate is the value of the kernel density estimation (KDE), which is shown by the green curves. The KDE uses Gaussian kernels where the standard deviation is set to the average of the asymmetric errors (see \citet{Silverman1986} for the KDE definition).
\label{fig:alpha_max_histogram}}
\end{figure*}

We show the distribution of the low energy spectral indexes, $\alpha$, in Figure~\ref{fig:alpha_max_histogram}, left panel. The histogram contains 153 spectra. The green curve is a smoothed version of the distribution, using the kernel density estimation (KDE), for which the errors are taken into account (see \citet{Silverman1986} for the KDE definition). We use the average of the asymmetric errors as the standard deviation of the Gaussian kernels in the KDE. This is a reasonable choice since the highest posterior density credible intervals of the $\alpha$ parameter is roughly symmetric around its mean values.  We compare the values of the low energy spectral index, $\alpha$ (considering $1\sigma$ lower limit) with the "synchrotron line of death" value \citep{Preece1998, Kaneko2006}, $\alpha = -2/3$, which is shown by the red dashed line in Figure~\ref{fig:alpha_max_histogram}. We find that 56$\%$ of the analyzed spectra (within a $1\sigma$ error) violate the criteria set by the "synchrotron line of death". 

Global parameter distributions, such as this $\alpha$-distribution, contain varying number of spectra from each individual burst and the spectral index typically vary between timebins of the same burst. Therefore, it is difficult to draw firm conclusions on the emission mechanism based on the entire sample, since there is a bias towards strong bursts with many timebins (e.g. GRB 120323; see the purple line in Figure \ref{fig:correlations_ALL}).

The best way to constrain the emission mechanism during a pulse/burst is therefore to select the timebin that contains the largest value of the spectral index $\alpha$ in each pulse/burst. The reason for this is that physical models typically have an upper limit on how hard the spectra can get. Therefore, it is enough that one single bin violates such a limit for the corresponding emission model to be rejected by the data, under the assumption that a single emission mechanism is the source of the observed signal in the entire duration of the burst  
\citep[e.g.,][]{Yu2019, Ryde2019, Acuner2019}. Some time-resolved spectral catalogs present the evolution of the parameters over all timebins \citep[e.g.,][]{Kaneko2006, Yu2016, Yu2019}. Their parameter relations for individual GRBs are then interpreted by physical models \citep[e.g., a qualitative photospheric emission scenario,][]{Ryde2019}.

\begin{figure*}
\centering
\subfigure{\includegraphics[align=t,width=0.45\linewidth]{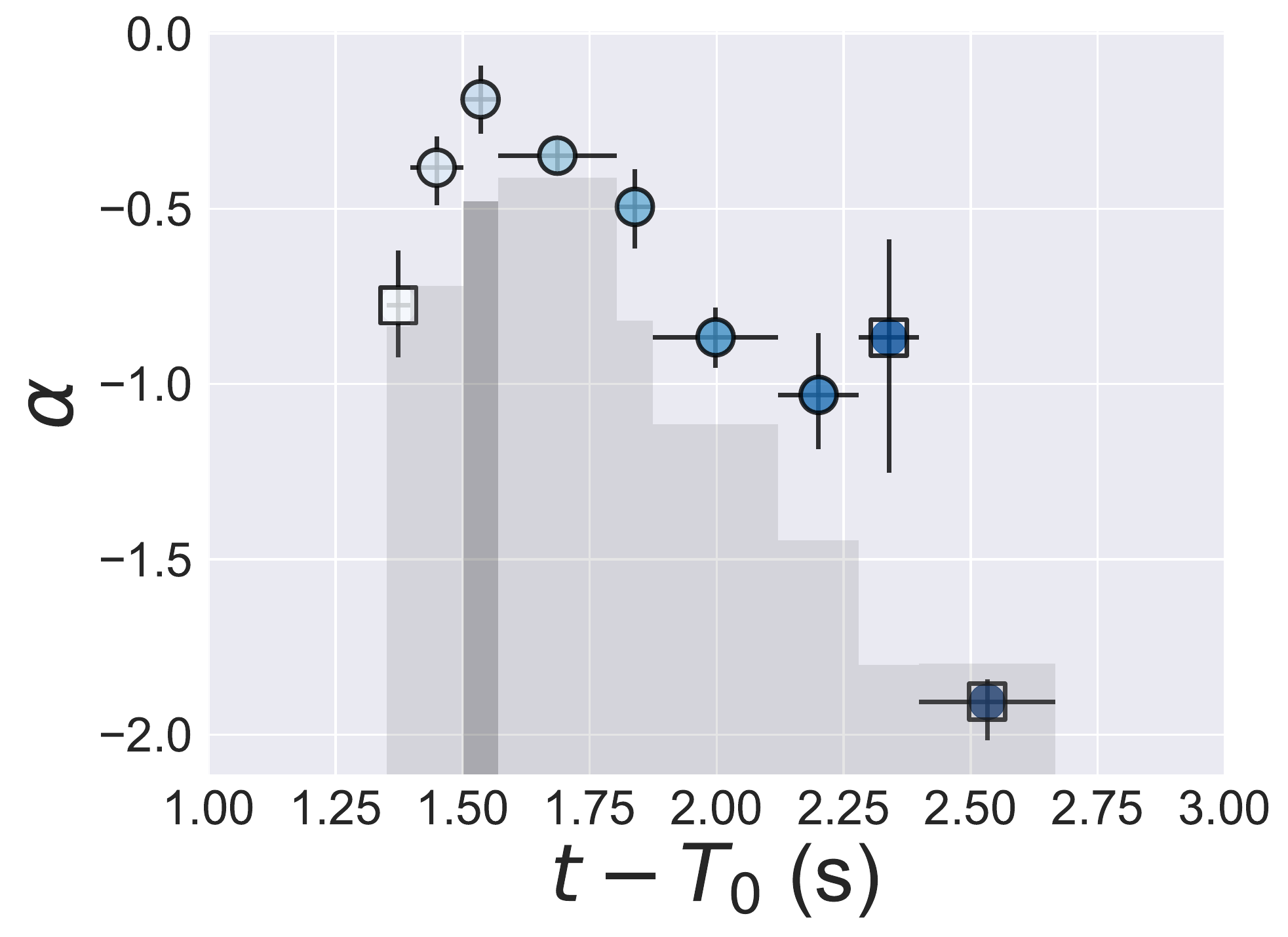}}
\subfigure{\includegraphics[align=t,width=0.45\linewidth]{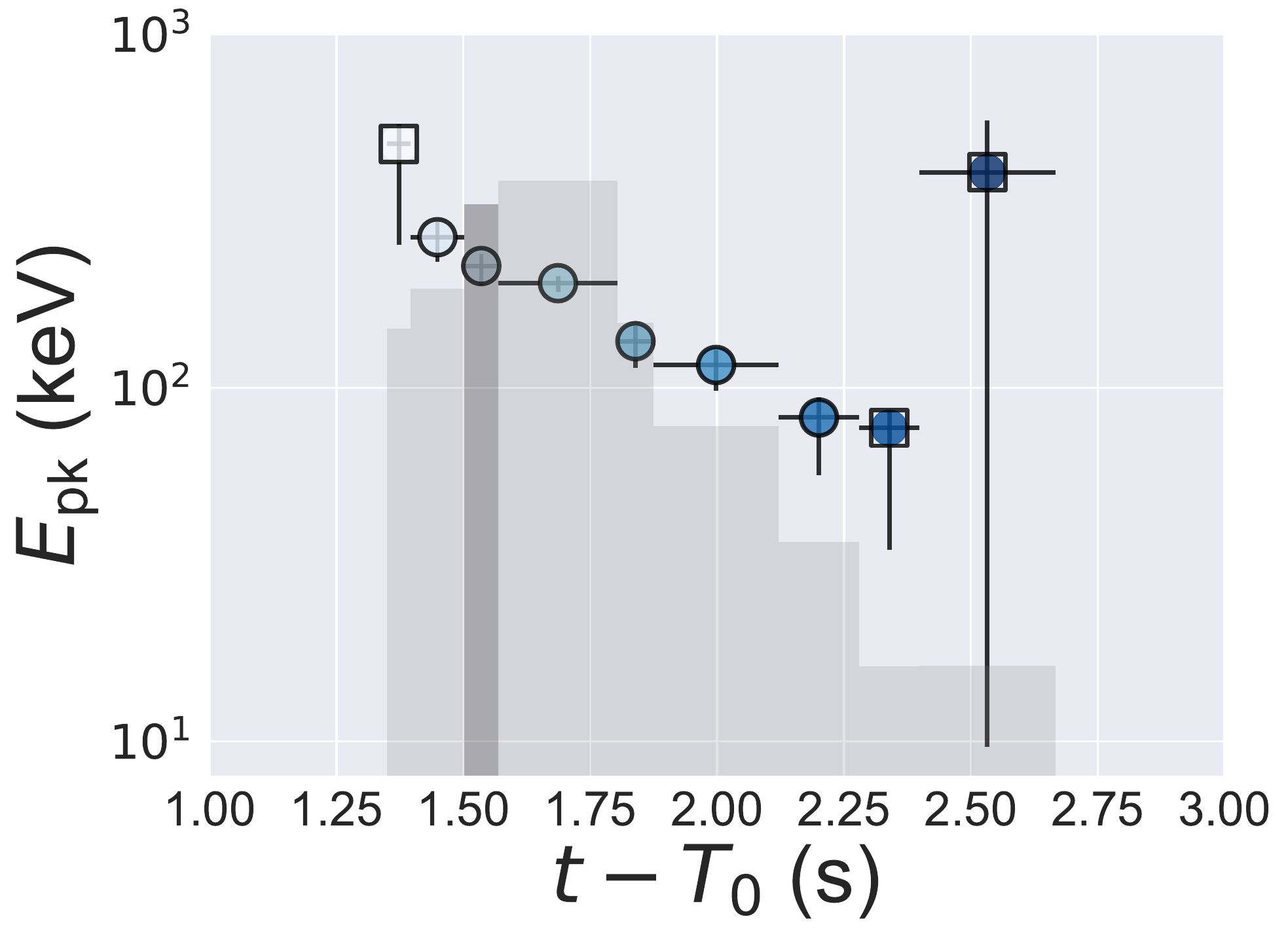}}
\caption{Spectral evolution of GRB140209 shown as an example. 
The spectral index  $\alpha$ (left panel) and  the peak energy $E_{\rm pk}$ (right panel) are shown together with the energy flux light curve (in arbitrary logarithmic units) overlaid in gray.
Only the timebins (from light to dark blue) with a statistical significance, $S\geq15$ are shown. Data points with circles indicate statistical significance $S\geq20$.
In left panel, it is seen that the spectral index $\alpha$ peaks close to the peak of the light curve. This is the typical behavior of most pulses \citep{Yu2019, Ryde2019}. The $\alpha_{\max}$ timebin is marked as dark grey.
\label{fig:example_LC}}
\end{figure*}

For each of the 70 pulses in the 68 short bursts in our sample we therefore select the timebin which contains the maximal (hardest) value of the low energy spectral index, which is denoted by $\alpha_{\rm max}$. The spectral index $\alpha_{\rm max}$ and its corresponding peak energy ($E_{\rm pk}$), flux and statistical significance ($S$) are listed in the last four Columns of Table \ref{tab:sample1}. We find that for most pulses, $\alpha_{\rm max}$ occurs close to, or at the peak of the light curve. As such, it contains the most valuable information of the spectra. This is demonstrated in GRB140209 shown in Figure \ref{fig:example_LC}: the temporal evolution of $\alpha$ and peak energy $E_{\rm pk}$ are shown, overlaid on the energy flux light curve in grey.

We now show in Figure~\ref{fig:alpha_max_histogram} (right panel) the distribution of $\alpha_{\rm max}$ in each of the 70 pulses from 68 short GRBs in the sample together with the KDE of the distribution. The red dashed line again shows the "synchrotron line of death", $\alpha = -2/3$. We find that 70$\%$ (within a $1\sigma$ error) of the pulses violate the "line of death" criterion, and are therefore better interpreted with a model that is not synchrotron, such as the photospheric model. This fraction is larger than the one obtained in the case of long GRBs, $60\%$ within a $1\sigma$ error \citep{Yu2019}. We also note that the softest value is $\alpha_{\rm max}=-1.26$ for GRB 160822.


\begin{deluxetable}{cccccccccccccc}
\tabletypesize{\scriptsize}
\tablecaption{A sample of 70 pulses from 68 short GRBs used in this study. Column 1: GRB names (bn), Column 2: Burst duration, Column 3: Detectors; the brightest one is in brackets and is used for background and Bayesian blocks determination. Column 4: Source interval, Columns 5 and 6: Background intervals, Column 7: Number of Bayesian blocks during the source interval. Column 8: Maximal low-energy spectral index. Column 9: Significance of the timebin with $\alpha_{\rm max}$. Column 10:
Corresponding peak energy ($E_{\rm pk}$), Column 11: Corresponding flux.
\label{tab:sample1}}
\tablehead{ 
\colhead{bn} & $T_{90}$ & \colhead{Detectors} & \colhead{$\Delta T_{\rm src}$} & \colhead{$\Delta T_{\rm bkg,1}$} & \colhead{$\Delta T_{\rm bkg,2}$}  & \colhead{$N$} & \colhead{$\alpha_{\rm max}$} & \colhead{$S$} & \colhead{$E_{\rm pk}$} & \colhead{Flux} \\
 \colhead{} & \colhead{(s)} & \colhead{} & \colhead{(s)} & \colhead{(s)} & \colhead{(s)} & \colhead{} & \colhead{} & \colhead{} & \colhead{(keV)} & \colhead{(${\rm~erg}{\rm~cm}^{-2}{\rm s}^{-1}$)} 
}
\startdata   
081209981 & 0.19$\pm$0.14& (n8)nbb1&$-1.7$ to 2. & $-20$. to $-7.$& 7. to 13.& 5 & $-0.42_{-0.13}^{+0.09}$ & 19 & $1080_{-250}^{+190}$ & $1.6_{-0.9}^{+2.2}\times10^{-5}$ \\
081216531 & 0.8$\pm$0.4 & n7(n8)nbb1  & $-1.7$ to 2. & $-20.$ to $-7.$  & 15. to 35. & 8 & $-0.37_{-0.07}^{+0.06}$ & 30 & $1170_{-110}^{+140}$ & $1.6_{-0.7}^{+1.1}\times10^{-5}$\\
081223419 & 0.58$\pm$0.14 & n6(n7)n9b1& $-2.$ to 2. &$-20.$ to $-7.$ &15. t0 35. & 4 & $-0.42_{-0.17}^{+0.13}$ & 18 &  $200_{-50}^{+30}$ & $2.0_{-1.2}^{+3.4}\times10^{-6}$ \\
090108020 & 0.71$\pm$0.14 & (n1)n2n5b0 & $-2.$ to 2. & $-20.$ to $-7.$  & 15. to 35. & 6 & $-0.22_{-0.15}^{+0.11}$ & 25 & $140_{-20}^{+10}$ & $2.6_{-1.3}^{+2.8}\times10^{-6}$ \\
090227772 & 0.31$\pm$0.02 & n0(n1)n2b0 & $-2.$ to 2. & $-20.$ to $-7.$ & 15. to 35. & 9 & $+0.07_{-0.07}^{+0.07}$ & 36 & $1720_{-140}^{+110}$ & $8.2_{-3.4}^{+5.7}\times10^{-5}$ \\
090228204 & 0.45$\pm$0.14 & (n0)n1n3b0 & $-2.$ to 2. & $-20.$ to $-7.$ & 15. to 35. & 11 & $-0.01_{-0.07}^{+0.06}$ & 45 & $640_{-50}^{+50}$ & $13_{-5.7}^{+10}\times10^{-5}$ \\
090308734 & 1.67$\pm$0.29 & (n3)n6n7b0 &$-2.$ to 2.5 & $-20.$ to $-7.$ & 15. to 35. & 6 & $-0.42_{-0.10}^{+0.09}$ & 19 &  $750_{-160}^{+100}$ & $3.0_{-1.6}^{+3.5}\times10^{-6}$ \\
090328713 & 0.2$\pm$1.0 &(n9)nanbb1 &$-1.$ to 5. &$-20.$ to $-7.$ & 15. to 35.& 3 & $-0.87_{-0.08}^{+0.06}$ & 17 &  $1700_{-570}^{+310}$ & $5.6_{-2.4}^{+3.6}\times10^{-6}$ \\
090510016 & 0.96$\pm$0.14 & (n6)n7n9b1 & $-1.7$ to 2. & $-20.$ to $-7.$ & 15. to 35. & 9 & $-0.62_{-0.05}^{+0.04}$ & 32 & $2850_{-340}^{+270}$ & $2.1_{-0.6}^{+1.0}\times10^{-5}$ \\
090617208 & 0.19$\pm$0.14 & n0(n1)n3b0 & $-1.$ to 2.5&$-20.$ to $-7.$ & 15. to 35. & 4 & $+0.05_{-0.19}^{+0.18}$ &  16 &  $920_{-310}^{+120}$ & $1.1_{-0.9}^{+3.7}\times10^{-5}$ \\
090802235 & 0.04$\pm$0.02 & n2(n5)b0 & $-3.$ to 8 & $-20.$ to $-7.$ & 20. to 40. & 4 &  $-0.48_{-0.14}^{+0.14}$ & 21 & $480_{-150}^{+60}$ & $1.4_{-0.9}^{+2.7}\times10^{-5}$ \\
090907808 & 0.8$\pm$0.3 & n6(n7)n9b1 &$-2.$ to 2. &$-20.$ to $-7.$ & 15. to 35.& 3 & $-0.09_{-0.16}^{+0.15}$ & 17 &  $450_{-100}^{+60}$ & $1.8_{-1.2}^{+4.4}\times10^{-6}$ \\
100223110 & 0.26$\pm$0.09 & n7(n8)b1& $-2.$ to 2.1 &$-20.$ to $-7.$ &15. to 35. & 5 & $-0.14_{-0.12}^{+0.11}$ & 20 &  $1250_{-260}^{+150}$ & $0.9_{-0.5}^{+1.4}\times10^{-5}$ \\
100629801 & 0.8$\pm$0.4 & (na)nbb1& $-2.$  to 2.& $-20.$ to $-7.$ &15. to 35. & 5 & $-0.76_{-0.18}^{+0.15}$ & 16 &  $230_{-80}^{+30}$ & $2.8_{-1.8}^{+5.6}\times10^{-6}$ \\
100811108 & 0.39$\pm$0.09 & (n7)n9nbb1& $-2.$ to 2.& $-20.$ to $-7.$ &15. to 35. & 3 & $-0.14_{-0.11}^{+0.09}$ &  19 &  $1140_{-220}^{+140}$ & $6.9_{-3.8}^{+8.2}\times10^{-6}$ \\
100827455 & 0.6$\pm$0.4 & n6(n7)n8b1 & $-2.$ to 2.& $-20.$ to $-7.$ &15. to 35. & 5 & $-0.32_{-0.11}^{+0.12}$ & 16 &  $780_{-170}^{+110}$ & $0.8_{-0.5}^{+1.1}\times10^{-5}$ \\
101216721 & 1.9$\pm$0.6 & n1n2(n5)b0 & $-2.$ to 3. & $-20.$ to $-7.$ & 15. to 35. & 8 & $-0.54_{-0.10}^{+0.08}$ & 37 & $180_{-20}^{+20}$ & $3.2_{-1.4}^{+2.5}\times10^{-6}$ \\
110212550 & 0.07$\pm$0.04 & n6(n7)n8b1 &$-1.8$ to 1.7 &$-20.$ to $-7.$ & 15. to 35.& 5 & $-0.38_{-0.14}^{+0.11}$ & 18 &  $620_{-150}^{+90}$ & $1.7_{-1.0}^{+2.9}\times10^{-5}$ \\
110526715 & 0.45$\pm$0.05 & (n3)n4b0 &$-1.7$ to 2. & $-20.$ to $-7.$ & 15. to 35.& 3 & $-0.88_{-0.10}^{+0.09}$ & 17 &  $670_{-240}^{+110}$ & $2.2_{-1.0}^{+1.8}\times10^{-6}$ \\
110529034 & 0.51$\pm$0.09 & n6n7(n9)b1 & $-2.$ to 2. & $-20.$ to $-7.$  & 15. to 35. & 10 & $-0.52_{-0.19}^{+0.15}$ & 15 & $780_{-350}^{+120}$ & $1.0_{-0.7}^{+2.4}\times10^{-5}$ \\
110705151 & 0.19$\pm$0.04 & (n3)n4n5b0 & $-1.7$ to 2. & $-20.$ to $-7.$  & 15. to 35. & 8 & $-0.09_{-0.12}^{+0.11}$ & 23 & $890_{-180}^{+90}$ & $1.6_{-1.0}^{+2.2}\times10^{-5}$ \\
111222619 & 0.29$\pm$0.04 & (n8)nbb1 & $-2.$ to 2. & $-20.$ to $-7.$  & 15. to 35. & 7 &  $-0.21_{-0.12}^{+0.11}$ & 27 & $700_{-110}^{+70}$ & $1.5_{-0.8}^{+1.9}\times10^{-5}$ \\
120222021 & 1.09$\pm$0.14 & n3n4(n5)b0 & $-1.7$ to 2. & $-20.$ to $-7.$  & 15. to 35. & 8 & $-0.31_{-0.26}^{+0.18}$ & 34 & $120_{-30}^{+20}$ & $1.7_{-1.2}^{+4.9}\times10^{-6}$ \\
120323507 & 0.39$\pm$0.04 & n0(n3)b0 & $-2.$ to 2. & $-20.$ to $-7.$  & 10. to 30. & 16 & $-0.73_{-0.14}^{+0.15}$ & 25 & $420_{-160}^{+40}$ & $3.1_{-1.9}^{+5.6}\times10^{-5}$ \\
120519721 & 1.1$\pm$0.5 & n7(n8)b1 & $-1.7$ to 2. & $-20.$ to $-7.$  & 15. to 35. & 6 & $-0.45_{-0.12}^{+0.09}$ & 23 & $860_{-190}^{+130}$ & $3.9_{-2.1}^{+5.5}\times10^{-6}$ \\
120624309 & 0.64$\pm$0.16 & n1n2(na)b0 & $-2.$ to 2. & $-20.$ to $-7.$  & 15. to 35. & 11 & $-0.68_{-0.05}^{+0.05}$ & 28 & $3590_{-540}^{+370}$ & $2.4_{-0.7}^{+1.0}\times10^{-5}$ \\
120811014 & 0.45$\pm$0.09 & n7(n8)b1 & $-2.$ to 3. & $-20.$ to $-7.$  & 15. to 35. & 5 &  $+0.01_{-0.16}^{+0.12}$ & 21 & $1170_{-230}^{+180}$ & $1.0_{-0.6}^{+2.0}\times10^{-5}$ \\
120817168 & 0.16$\pm$0.11 & n6(n7)n8b1 & $-1.5$ to 1.7 & $-20.$ to $-7.$  & 15. to 35. & 6 & $-0.50_{-0.07}^{+0.06}$ & 30 & $1530_{-290}^{+190}$ & $3.9_{-1.5}^{+2.4}\times10^{-5}$ \\
120830297 & 0.90$\pm$0.23 & (n0)n1n3b0 & $-1.7$ to 2. & $-20.$ to $-7.$  & 15. to 35. & 5 & $-0.27_{-0.10}^{+0.07}$ & 24 & $1100_{-180}^{+150}$ & $3.6_{-1.7}^{+3.2}\times10^{-6}$ \\
121127914 & 0.6$\pm$0.4 & (n4)n8b1& $-2.$ to 2.& $-20.$ to $-7.$ &15. to 35. & 4 & $-0.53_{-0.11}^{+0.10}$ & 19 &  $1080_{-270}^{+170}$ & $6.6_{-3.6}^{+8.8}\times10^{-6}$ \\
130416770 & 0.2$\pm$0.4 & n3(n4)n5b0 & $-1.7$ to 8.& $-20.$ to $-7.$ &25. to 45. & 3 & $-0.52_{-0.09}^{+0.08}$ & 16 &  $1100_{-230}^{+170}$ & $1.0_{-0.5}^{+1.0}\times10^{-5}$ \\
130504314 & 0.39$\pm$0.18 & (n3)n4b0 & $-2.$ to 1.7 & $-35.$ to $-7.$  & 15. to 30. & 7 & $-0.09_{-0.11}^{+0.08}$ & 21 & $1370_{-190}^{+150}$ & $1.4_{-0.7}^{+1.7}\times10^{-5}$ \\
130628860 & 0.51$\pm$0.14 & n7n9(nb)b1 &$-2.$ to 2.1 &$-20.$ to $-7.$ &15. to 35. & 7 & $-0.18_{-0.13}^{+0.14}$ & 18 &  $1120_{-270}^{+150}$ & $1.8_{-1.2}^{+3.3}\times10^{-5}$ \\
130701761 & 1.60$\pm$0.14 &(n9)nanbb1 &$-2.$ to 3. & $-20.$ to $-7.$ &15. to 35. & 8 & $-0.24_{-0.05}^{+0.18}$ & 18 &  $1200_{-330}^{+30}$ & $4.4_{-3.0}^{+8.6}\times10^{-6}$ \\
130804023 & 0.96$\pm$0.09 & n6(n7)n9b1 & $-2.$ to 2. & $-8.$ to $-5.$  & 12. to 29. & 10 & $-0.26_{-0.10}^{+0.10}$ & 20 & $850_{-150}^{+109}$ & $3.1_{-1.7}^{+3.6}\times10^{-5}$ \\ 
130912358 & 0.51$\pm$0.14 & n7(n8)nbb1 & $-2.$ to 2.& $-20.$ to $-7.$ &15. to 35. & 8 & $-1.02_{-0.09}^{+0.08}$ & 18 &  $1700_{-870}^{+230}$ & $7.2_{-3.4}^{+6.6}\times10^{-6}$ \\
131126163 & 0.2$\pm$0.4 & n2(n5)b0&$-3$ to 1.7 & $-25.$ to $-10.$&15. to 35. & 4 & $-0.07_{-0.16}^{+0.15}$ & 19 &  $750_{-190}^{+100}$ & $1.9_{-1.3}^{+4.3}\times10^{-5}$ \\
140209313 & 1.41$\pm$0.27 & n9(na)b1 & 0. to 4. & $-20.$ to $-7.$  & 20. to 40. & 13 &  $-0.19_{-0.10}^{+0.10}$ & 44 & $220_{-30}^{+20}$ & $1.4_{-0.7}^{+1.4}\times10^{-5}$ \\
140807500 & 0.5$\pm$0.2 & n3(n4)n5b0& $-2.$ to 2. & $-20.$ to $-7.$  & 15. to 35. & 4 & $-0.75_{-0.09}^{+0.07}$ & 26 & $750_{-190}^{+110}$ & $5.2_{-2.2}^{+4.0}\times10^{-6}$ \\
140901821 & 0.18$\pm$0.04 & n9(na)nbb1 & $-1.5$ to 2. & $-20.$ to $-7.$  & 15. to 35. & 5 & $-0.22_{-0.06}^{+0.05}$ & 36 & $1200_{-100}^{+90}$ & $2.5_{-0.9}^{+1.3}\times10^{-5}$ \\
\enddata
\end{deluxetable}


\begin{deluxetable}{cccccccccccccc}
\tabletypesize{\scriptsize}
\tablecaption{Table \ref{tab:sample1} (\textit{continued}) \label{tab:sample2}}
\tablehead{ 
\colhead{bn} & $T_{90}$ & \colhead{Detectors} & \colhead{$\Delta T_{\rm src}$} & \colhead{$\Delta T_{\rm bkg,1}$} & \colhead{$\Delta T_{\rm bkg,2}$}  & \colhead{$N$} & \colhead{$\alpha_{\rm max}$} & \colhead{$S$} & \colhead{$E_{\rm pk}$} & \colhead{Flux} \\
 \colhead{} & \colhead{(s)} & \colhead{} & \colhead{(s)} & \colhead{(s)} & \colhead{(s)} & \colhead{} & \colhead{} & \colhead{} & \colhead{(keV)} & \colhead{ $({\rm~erg}{\rm~cm}^{-2}{\rm s}^{-1}$)} 
}
\startdata
141011282 & 0.08$\pm$0.04 & n0(n1)n9b0 & $-1.8$ to 1.7 & $-20.$ to $-7.$  & 15. to 35. & 5 & $-0.46_{-0.11}^{+0.10}$ & 20 & $790_{-180}^{+100}$ & $2.7_{-1.4}^{+3.4}\times10^{-5}$ \\ 
141105406 & 1.28$\pm$1.03 & n6n7(n9)b1&$-1.7$ to 2. &$-20.$ to $-7.$ &15. to 35. & 4 & $-0.42_{-0.13}^{+0.11}$ & 18 &  $500_{-120}^{+70}$ & $2.2_{-1.4}^{+3.4}\times10^{-6}$ \\
141202470 & 1.41$\pm$0.27 & (n7)n8nbb1 & $-1.7$ to 3. & $-20.$ to $-7.$  & 15. to 35. & 4 & $-0.32_{-0.10}^{+0.08}$  & 20 & $840_{-150}^{+110}$ & $3.5_{-1.7}^{+3.6}\times10^{-6}$ \\ 
141213300 & 0.8$\pm$0.5 & n1(n2)n5b0&$-1.7$ to 2. &$-20.$ to $-7.$ &15. to 35. & 7 & $-0.87_{-0.16}^{+0.14}$ & 15 &  $210_{-90}^{+30}$ & $1.7_{-1.0}^{+2.9}\times10^{-6}$ \\
150118927 & 0.3$\pm$0.1 & n7n8(nb)b1 & $-1.7$ to 1.7 & $-20.$ to $-7.$  & 15. to 35. & 6 & $-0.65_{-0.10}^{+0.08}$ & 26 & $	620_{-160}^{+90}$ & $1.4_{-0.7}^{+1.4}\times10^{-5}$ \\ 
150810485 & 1.3$\pm$1.0 & n6n7(nb)b1&$-1.7$ to 2. &$-20.$ to $-7$ &15. to 35. & 6& $-0.61_{-0.08}^{+0.07}$ & 15 &  $1530_{-360}^{+250}$ & $5.4_{-2.5}^{+4.4}\times10^{-6}$ \\
150811849 & 0.64$\pm$0.14 &(n4)n5b0 & $-2.$ to 2.& $-20.$ to $-7.$ &15. to 35. &6 & $-0.24_{-0.11}^{+0.08}$ & 18 &  $1600_{-270}^{+220}$ & $8.3_{-4.2}^{+9.3}\times10^{-6}$ \\
150819440 & 0.96$\pm$0.09 & n2(na)b1 & $-0.2$ to 0.3 & $-20.$ to $-7.$  & 15. to 35. & 9 & $+0.25_{-0.17}^{+0.14}$ & 33 & $440_{-70}^{+50}$ & $6.2_{-4.1}^{+12}\times10^{-5}$ \\
150819440 & 0.96$\pm$0.09 & n2(na)b1 & $0.35$ to 1.37 & $-20.$ to $-7.$  & 15. to 35. & 11 & $-1.02_{-0.05}^{+0.04}$ & 71 & $330_{-40}^{+30}$ & $2.0_{-0.5}^{+0.7}\times10^{-5}$ \\
150922234 & 0.15$\pm$0.04 &n6(n9)nab1 &$-1.5$ to 2. &$-20.$ to $-7.$ &15. to 35. & 7 & $-0.24_{-0.24}^{+0.16}$ & 15 &  $800_{-270}^{+150}$ & $1.1_{-0.9}^{+4.3}\times10^{-5}$ \\
150923864 & 1.79$\pm$0.09 & n6(n7)n9b1 &$-2.$ to 3.&$-20.$ to $-7.$ &15. to 35. & 9 & $+0.06_{-0.21}^{+0.25}$ & 15 &  $140_{-30}^{+20}$ & $1.5_{-1.1}^{+4.0}\times10^{-6}$ \\
151222340 & 0.8$\pm$0.4 & (n4)b0 &$-1.7$ to 2. &$-20.$ to $-7.$ &15. to 35. & 6 & $-0.43_{-0.12}^{+0.09}$ & 16 &  $1500_{-320}^{+200}$ & $4.6_{-2.5}^{+6.6}\times10^{-6}$ \\
151231568 & 0.8$\pm$0.4 & n6(n7)n8b1& $-1.7$ to 2.&$-20.$ to $-7.$ &15. to 35. & 6 & $-0.58_{-0.11}^{+0.10}$ & 19 &  $610_{-180}^{+70}$ & $4.1_{-2.3}^{+4.7}\times10^{-6}$ \\
160408268 & 1.1$\pm$0.6 & (n0)n1n3b0&$-1.7$ to 2. &$-20.$ to $-7.$ &15. to 35. & 3 & $-0.77_{-0.11}^{+0.08}$ & 17 &  $1070_{-380}^{+180}$ & $3.0_{-1.5}^{+3.5}\times10^{-6}$ \\
160612842 & 0.29$\pm$0.23 & n0(n1)b0 &$-1.7$ to 1.7 &$-20.$ to $-7.$ &15. to 35. & 4 & $-0.78_{-0.12}^{+0.08}$ & 15 &  $2150_{-880}^{+510}$ & $5.4_{-2.6}^{+6.4}\times10^{-6}$ \\
160726065 & 0.8$\pm$0.4 & n0(n1)n2b0 & $-1.7$ to 2. & $-20.$ to $-7.$  & 15. to 35. & 6 & $-0.79_{-0.11}^{+0.09}$ & 21 & $460_{-140}^{+70}$ & $4.4_{-2.0}^{+4.3}\times10^{-6}$ \\
160806584 & 1.7$\pm$0.5 & (n8)nbb1& $-2.$ to 2.&$-20.$ to $-7.$ &15. to 35. & 6 & $-0.41_{-0.20}^{+0.14}$ & 19 &  $190_{-50}^{+30}$ & $2.1_{-1.4}^{+4.8}\times10^{-6}$ \\
160822672 & 0.1$\pm$0.4 & n9(na)b1 & $-2.$ to 1.7 & $-20.$ to $-7.$  & 15. to 35. & 5 & $-1.26_{-0.06}^{+0.03}$ & 23 & $290_{-50}^{+40}$ & $12_{-3.0}^{+4.5}\times10^{-5}$ \\
170127067 & 0.13$\pm$0.05 & (n4)b0 & $-2.$ to 2. & $-35.$ to $-15.$  & 35. to 55. & 5 & $+0.43_{-0.11}^{+0.16}$ & 32 & $900_{-90}^{+70}$ & $5.1_{-3.5}^{+12}\times10^{-5}$ \\
170206453 & 1.17$\pm$0.10 & (n9)nanbb1 & $-1.7$ to 3. & $-20.$ to $-7.$  & 15. to 35. & 9 & $-0.13_{-0.09}^{+0.07}$ & 39 & $370_{-40}^{+30}$ & $1.5_{-0.6}^{+1.2}\times10^{-5}$ \\
170222209 & 1.67$\pm$0.14 & n2(n5)b0 &$-1.7$ to 3. &$-20.$ to $-7.$ &20. to 40. & 10 & $-1.11_{-0.13}^{+0.10}$ & 15 &  $1550_{-1120}^{+130}$ & $5.8_{-3.3}^{+7.5}\times10^{-6}$ \\
170305256 & 0.45$\pm$0.07 & n0(n1)n2b0 & $-1.7$ to 2. & $-20.$ to $-7.$  & 15. to 35. & 6 & $-0.23_{-0.10}^{+0.11}$ & 25 & $340_{-50}^{+40}$ & $7.9_{-4.3}^{+8.0}\times10^{-6}$ \\
170708046 & 0.15$\pm$0.05 & n7(n8)nbb1 & $-2$ to 1.7 & $-20.$ to $-7.$  & 15. to 35. & 5 & $-0.75_{-0.10}^{+0.09}$ & 23 & $310_{-70}^{+40}$ & $1.2_{-0.6}^{+0.9}\times10^{-5}$ \\
170816599 & 1.60$\pm$0.14 & n7(n8)nbb1 & $-2.$ to 3. & $-20.$ to $-7.$  & 15. to 35. & 5 & $-0.26_{-0.07}^{+0.06}$ & 26 & $1170_{-120}^{+100}$ & $1.1_{-0.4}^{+0.6}\times10^{-5}$ \\
171108656 & 0.03$\pm$0.02 & (na)nbb1 & $-5.$ to 1.7 & $-30.$ to $-10.$  & 10. to 30. & 8 & $	-0.11_{-0.18}^{+0.16}$ & 25 & $120_{-20}^{+10}$ & $9.7_{-6.3}^{+16}\times10^{-6}$ \\
171126235 & 1.47$\pm$0.14 & n0(n1)n5b0 & $-1.7$ to 3. & $-20.$ to $-7.$  & 15. to 35. & 9 & $+0.09_{-0.17}^{+0.14}$ & 29 & $130_{-20}^{+10}$ & $4.4_{-2.5}^{+6.2}\times10^{-6}$ \\
180204109 & 1.15$\pm$0.09 & n3(n4)n5b0 & $-2.$ to 2.&$-20.$ to $-7.$ &15. to 35. & 12 & $-0.75_{-0.12}^{+0.12}$ & 16 &  $1300_{-660}^{+180}$ & $0.8_{-0.4}^{+1.1}\times10^{-5}$ \\
180703949 & 1.54$\pm$0.09 & n0n1(n3)b0 & $-0.30$ to 0.68 & $-20.$ to $-7.$  & 20. to 40. & 8 & $-0.35_{-0.12}^{+ 0.10}$ & 39 & $100_{-10}^{+10}$ & $4.7_{-2.2}^{+3.6}\times10^{-6}$ \\
180703949 & 1.54$\pm$0.09 & n0n1(n3)b0 & $0.7$ to 2.5 & $-20.$ to $-7.$  & 20. to 40. & 12 & $-0.24_{-0.05}^{+0.06}$ & 8 & $180_{-10}^{+10}$ & $1.5_{-0.4}^{+0.6}\times10^{-5}$ \\
180715741 & 1.7$\pm$1.4 & n3(n4)b0 &$-2.$ to 3. &$-30.$ to $-7.$ &20. to 45. & 5 & $-0.31_{-0.16}^{+0.16}$ & 16 &  $630_{-220}^{+90}$ & $1.8_{-1.2}^{+4.0}\times10^{-6}$ \\
\enddata
\end{deluxetable}


\subsection{On the consistency with non-dissipative photospheric model}
\label{subsect:NDP}

\begin{figure}
 \includegraphics[align=t,width=\columnwidth]{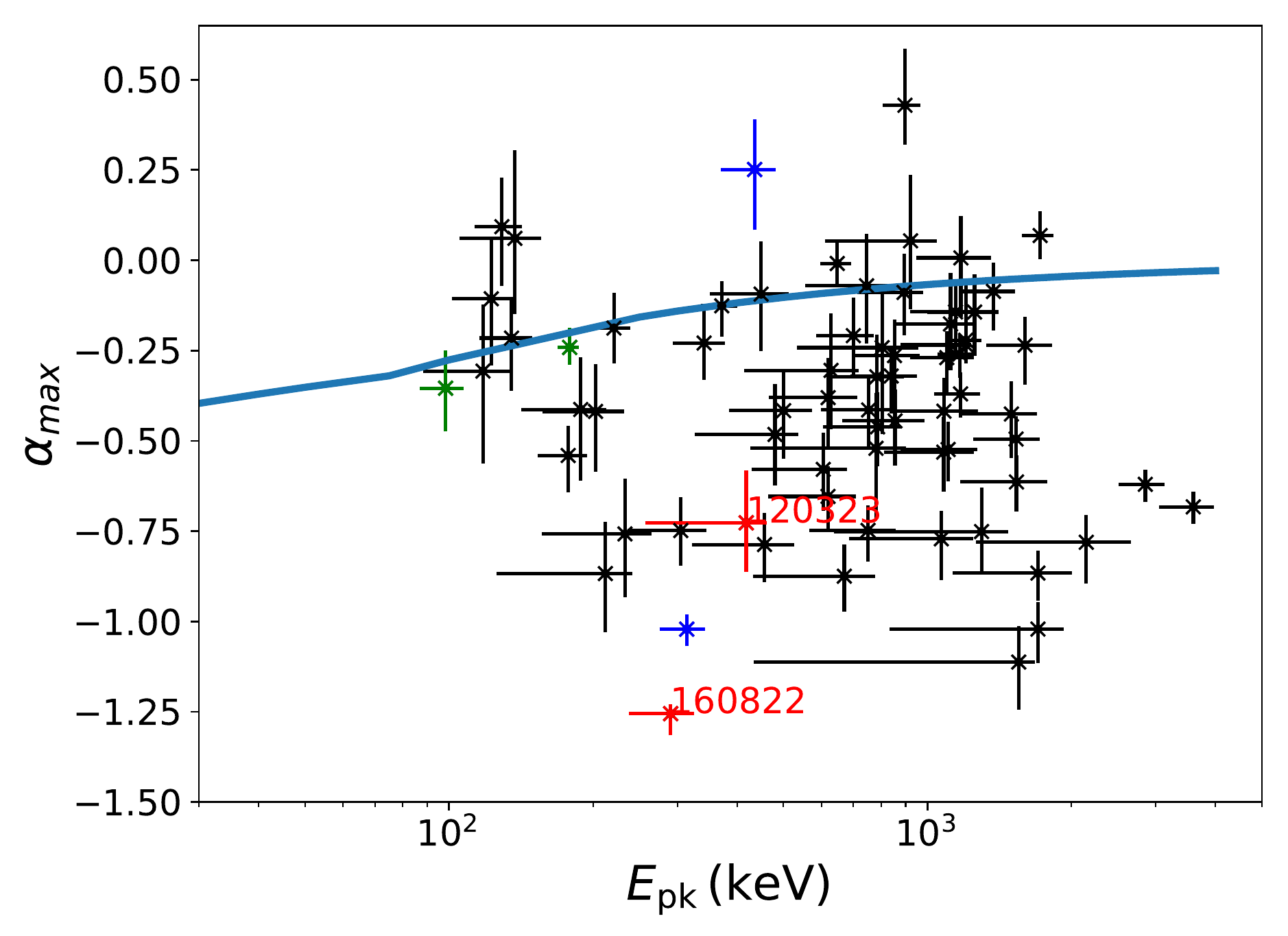}
 \caption{Relation of $\alpha_{\rm max}$ and $E_{\rm pk}$ for 70 pulses from 68 short GRBs. The light blue line is the expectation from a non-dissipative photosphere (NDP) spectrum (see Subsection \ref{subsect:NDP} for the definition). The fraction of pulses with $\alpha_{\rm max}$ larger than the NDP model prediction is $36\%$ (25 out of 70 pulses) within $1\sigma$ error. The blue and green points are for GRB 150819 and GRB 180703, which both have two pulses. GRB 120323 (which is the brightest short GRB) and GRB 160822 (which has the softest value of $\alpha_{\rm max}$) are shown in red.}
 \label{fig:Epk_vs_alpha_max}
\end{figure}

We show in Figure~\ref{fig:Epk_vs_alpha_max} the relation of $\alpha_{\rm max}$ and $E_{\rm pk}$ in all 70 pulses from 68 short GRBs in the sample. The light blue line corresponds to the values of $\alpha$ that are found when a CPL function is fitted to synthetic data, generated by a non-dissipative photosphere (NDP) spectrum  peaking at different energies, $E_{\rm pk}$, as described in \citet{Acuner2019} and shown in their Fig. 3. These $\alpha$-values (light blue line) are significantly smaller than the asymptotic slope of the theoretical NDP spectrum \citep[$\alpha\sim 0.4$,][]{Beloborodov2010, Begue2013, Ito2013, Lundman2013} due to (i) the limited energy band of the detector as well as (ii) the limitation of the CPL function to correctly model the shape of the true spectrum. The shape of the non-dissipative photosphere is shown in Fig. 1 in \citet{Ryde2017} and an analytical approximation is given in Equation (1) in \citet{Acuner2019}. 

From Figure~\ref{fig:Epk_vs_alpha_max}, we find that 36$\%$ of the observed points (25/70) are consistent with being above the NDP line within $1\sigma$ error, and are therefore consistent with having a dominant quasi black body component. This fraction is significantly larger than the fraction found in the study of pulses from long GRBs, 26$\%$ (for single or multi pulses bursts) and 28$\%$ (for only single pulsed bursts) from both catalogs of \citet{Yu2016} and \citet{Yu2019}, respectively, see \citet{Acuner2019} for further details. 

Even though most of the short GRBs in the sample are single pulse bursts, two bursts in our sample (GRB 150819 and GRB 180703) are found to have two separate pulses (these are marked by blue and green points respectively in Figure \ref{fig:Epk_vs_alpha_max}). While the first pulse of GRB 150819 (blue color in Figure \ref{fig:Epk_vs_alpha_max}) shows a hard spectral index, $\alpha_{\rm max}=0.25_{-0.17}^{+0.14}$, the spectral slope of the second pulse is much softer, $\alpha_{\rm max}=-1.02_{-0.05}^{+0.04}$. This might be an indication for a change in the leading emission mechanism, e.g., from photospheric emission to synchrotron \citep{Zhang2018}. On the other hand, both pulses of GRB 180703 (green color in Figure \ref{fig:Epk_vs_alpha_max}) are very hard, and are both compatible with NDP line. This might be an example of a burst in which a single emission mechanism is responsible for the full duration of a burst. In this case, the dominant emission mechanism throughout the full duration of the burst is likely a photospheric emission \citep{Acuner2018}.

\subsection{Lorentz factor}
\label{subsec:LF}
If indeed the observed spectra above the NDP line have a photospheric origin, then one can use the data to calculate the coasting Lorentz factor, $\eta$.  Here we estimate the Lorentz factor, $\eta$, for 25 pulses from 24 short GRBs above the NDP line by using equations (1) and (4) in \citet{Peer2007}. As the redshift of most bursts in our sample are unknown, we have assumed redshift $z=1$. We further assume the flux, $F$, in the analyzed timebin is equal to the black body flux, $\sim F_{\rm BB}$ and the observed temperature is related to the peak energy via $E_{\rm pk}\sim1.48T^{\rm obs}$. The flux and peak energy ($E_{\rm pk}$) for each short GRB obtained in our analysis for the corresponding $\alpha_{\rm max}$ timebin  are presented in Table~\ref{tab:sample1}. For comparison, we also compute the $\eta$ for all 12 long GRBs found above the NDP line in the sample of \citet{Yu2019, Acuner2019}. 

\begin{figure*}
\centering
\subfigure{\includegraphics[align=t,width=0.45\textwidth]{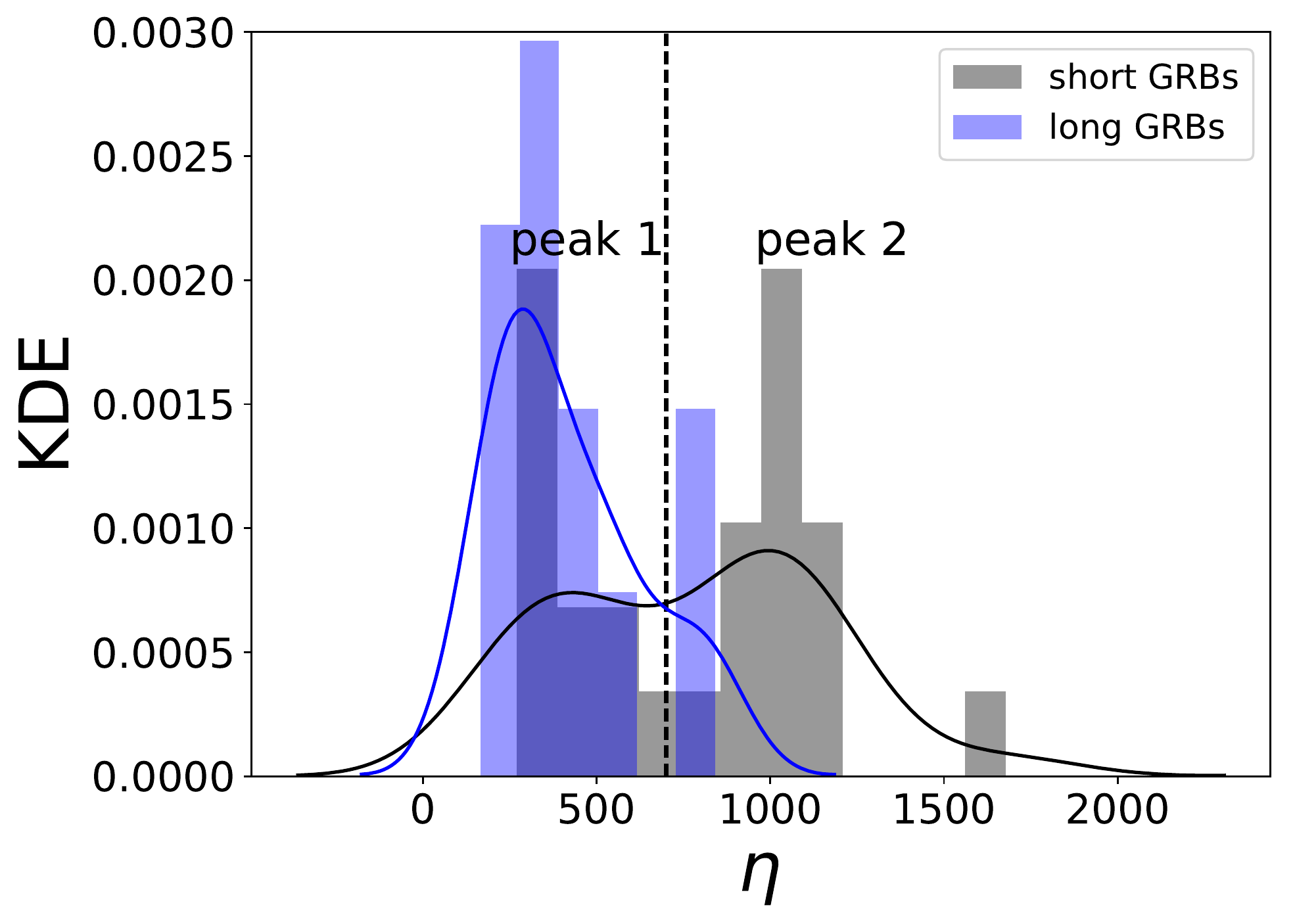}}
\subfigure{\includegraphics[align=t,width=0.45\textwidth]{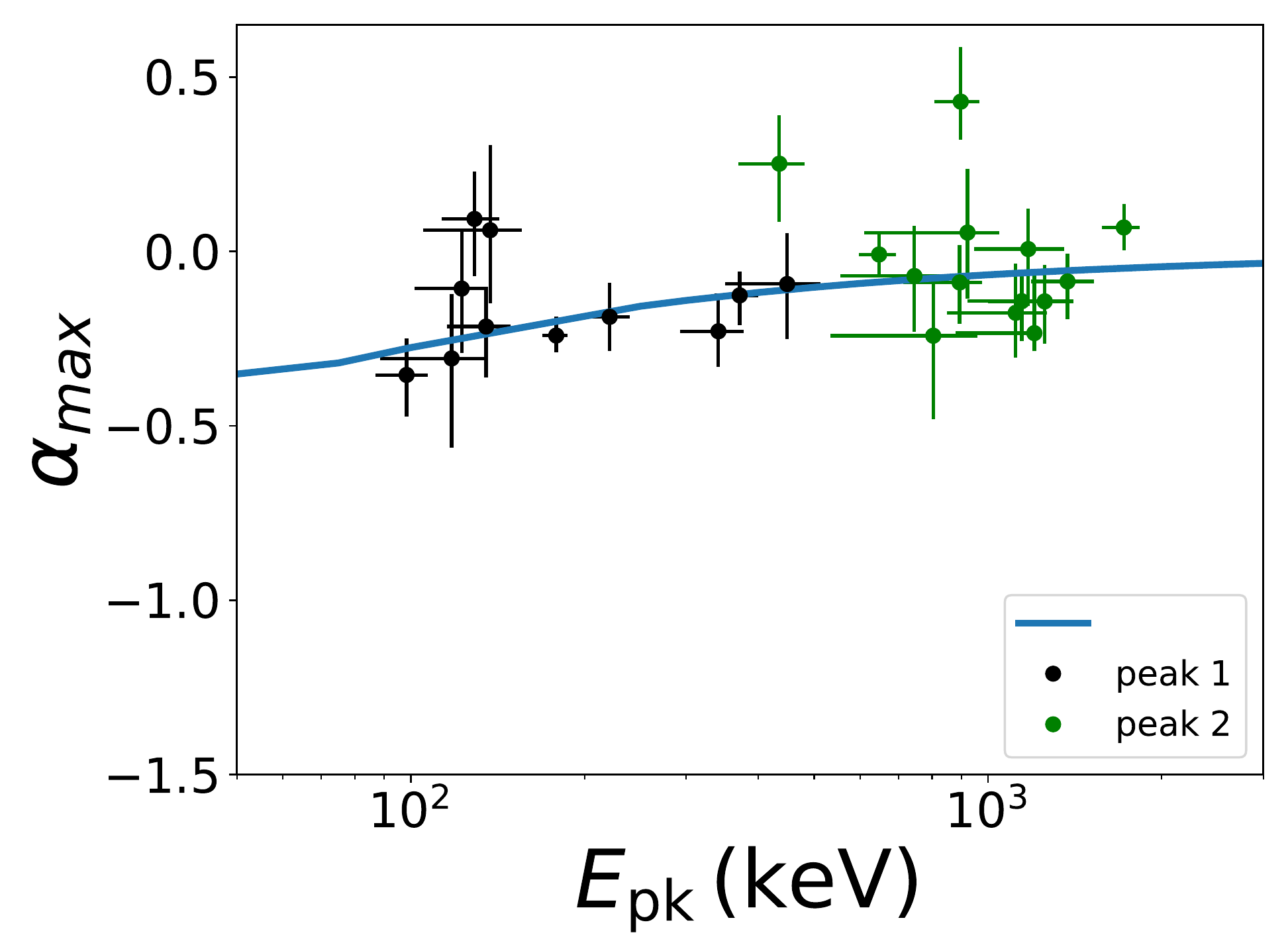}}
\caption{Left: Distributions of the Lorentz factor, $\eta$, for 25 pulses from 24 short GRBs (black histogram) and 12 long GRBs (blue historgram) above the NDP line. The curves represent the Kernel Density Estimation (KDE) of the distributions. The black dashed line indicate a cut at $\eta_c = 700$ to show the transition between peak 1 and peak 2 in the bi-modal distribution of short GRBs. Right: $\alpha_{\rm max}$ and $E_{\rm pk}$ relation (the  Spearman's  rank correlation coefficient is $r = 0.29$, the chance probability is $p = 0.16$). Data points are the short GRBs  from the bi-modal $\eta$ distribution (peak 1: black color and peak 2: green color). The light blue line (NDP line) is the same as in Figure \ref{fig:Epk_vs_alpha_max}. 
\label{fig:LFdistribution}}
\end{figure*}

The distributions of the Lorentz factor $\eta$ for 25 pulses from 24 short GRBs and 12 long GRBs above the NDP line are presented in Figure \ref{fig:LFdistribution}. We find that the mean Lorentz factor ($\eta_{\rm mean} = 775$) of the short GRBs 
is similar, though somewhat higher, than that of the long GRBs ($\eta_{\rm mean} = 416$) \citep{Peer2007, Racusin2011, Ghirlanda2018, Chen2018}. 

Surprisingly, we find a bi-modal distribution in the values of $\eta$ for short GRBs (Figure \ref{fig:LFdistribution}, left panel). There are 11 objects in "peak 1" and 14 in  "peak 2". However, such a bi-modal distribution is not found in the analysis of long GRBs. The low peak coincides with the values obtained for long GRBs, while the high peak is a factor of $\sim3$ higher.  When cutting the sample at $\eta_c = 700$, we find that the pulses with the high Lorentz factors, defined as "peak 2", have a corresponding higher $E_{\rm pk}$ (Figure \ref{fig:LFdistribution}, right panel). While this by itself may not be surprising, we point out that no such clear correlation is observed when analyzing the entire population of 70 pulses (see Figure \ref{fig:Epk_vs_alpha_max}).  

To validate the existence of this bimodality, we applied the dip test \citep{Hartigan&Hartigan1985} in the R package\footnote{https://github.com/alimuldal/diptest} to the sample of 25 pulses. The dip test results in 0.092, implying that the Lorentz factor distribution is bi-modal at a confidence level of 95\%.

\subsection{Correlations in bursts above the NDP line}
\label{subsec:correlations}

\begin{figure*}
\centering
\subfigure{\includegraphics[align=t,width=0.45\textwidth]{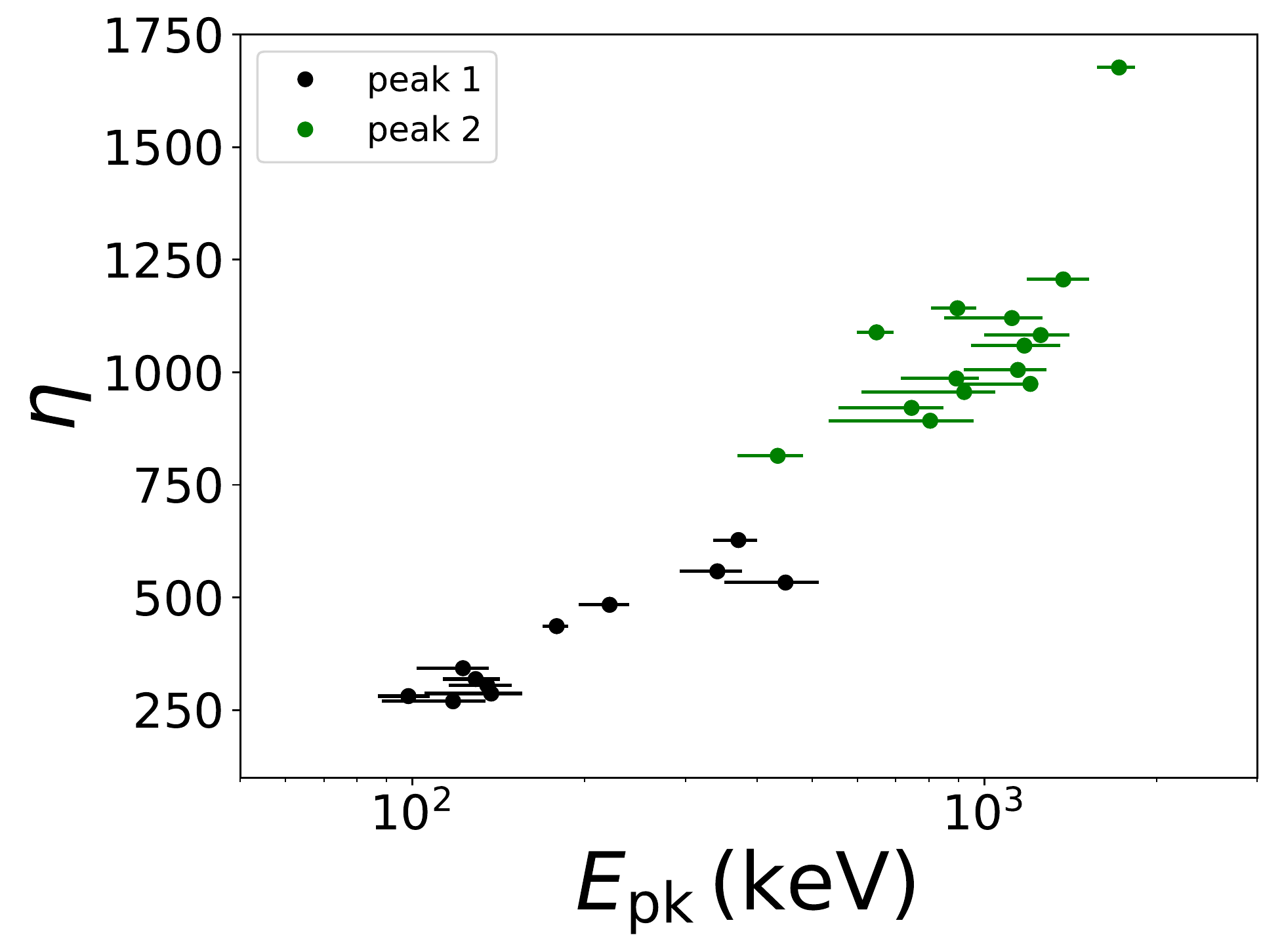}}
\subfigure{\includegraphics[align=t,width=0.45\textwidth]{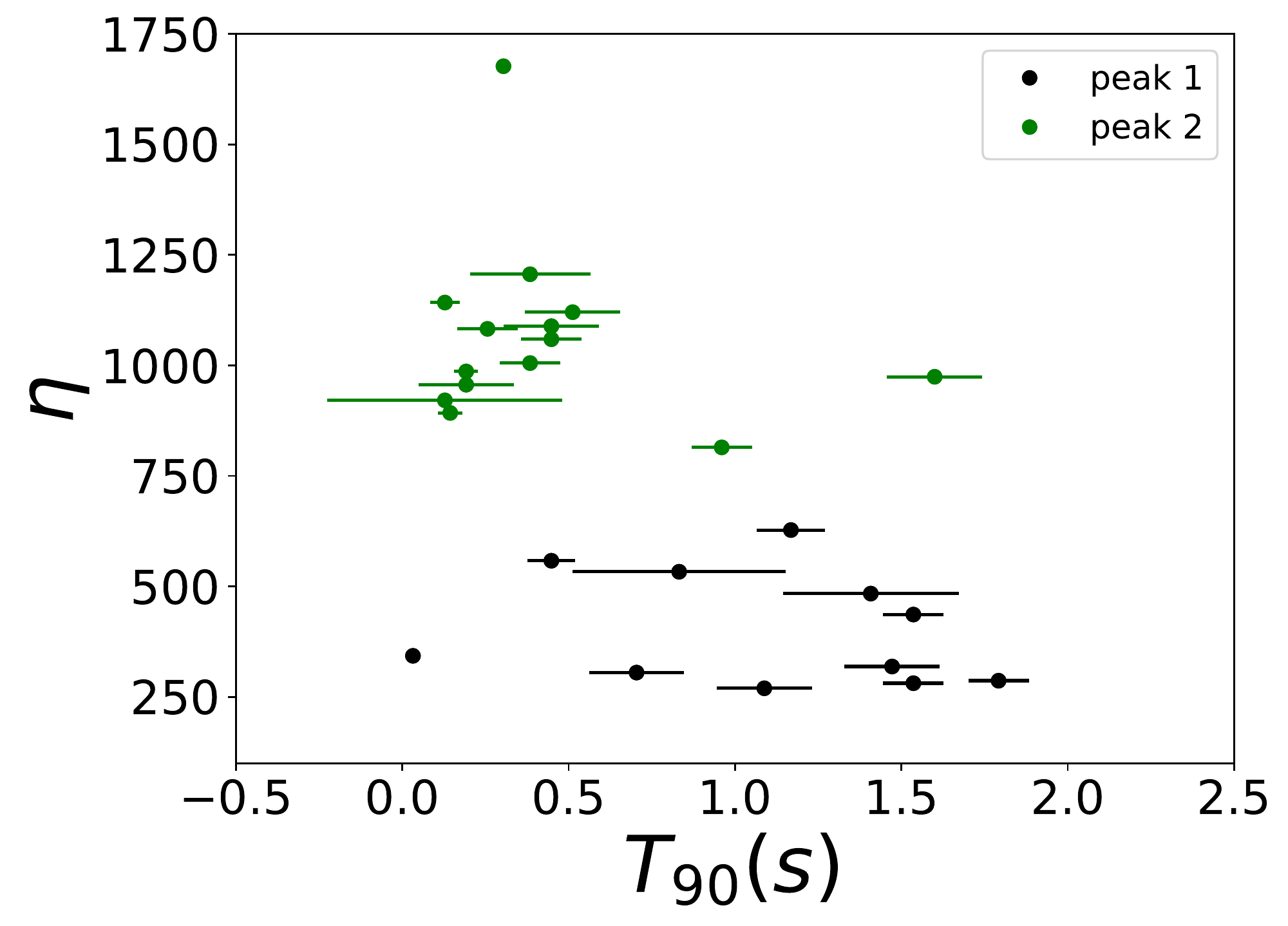}}
\subfigure{\includegraphics[align=t,width=0.45\textwidth]{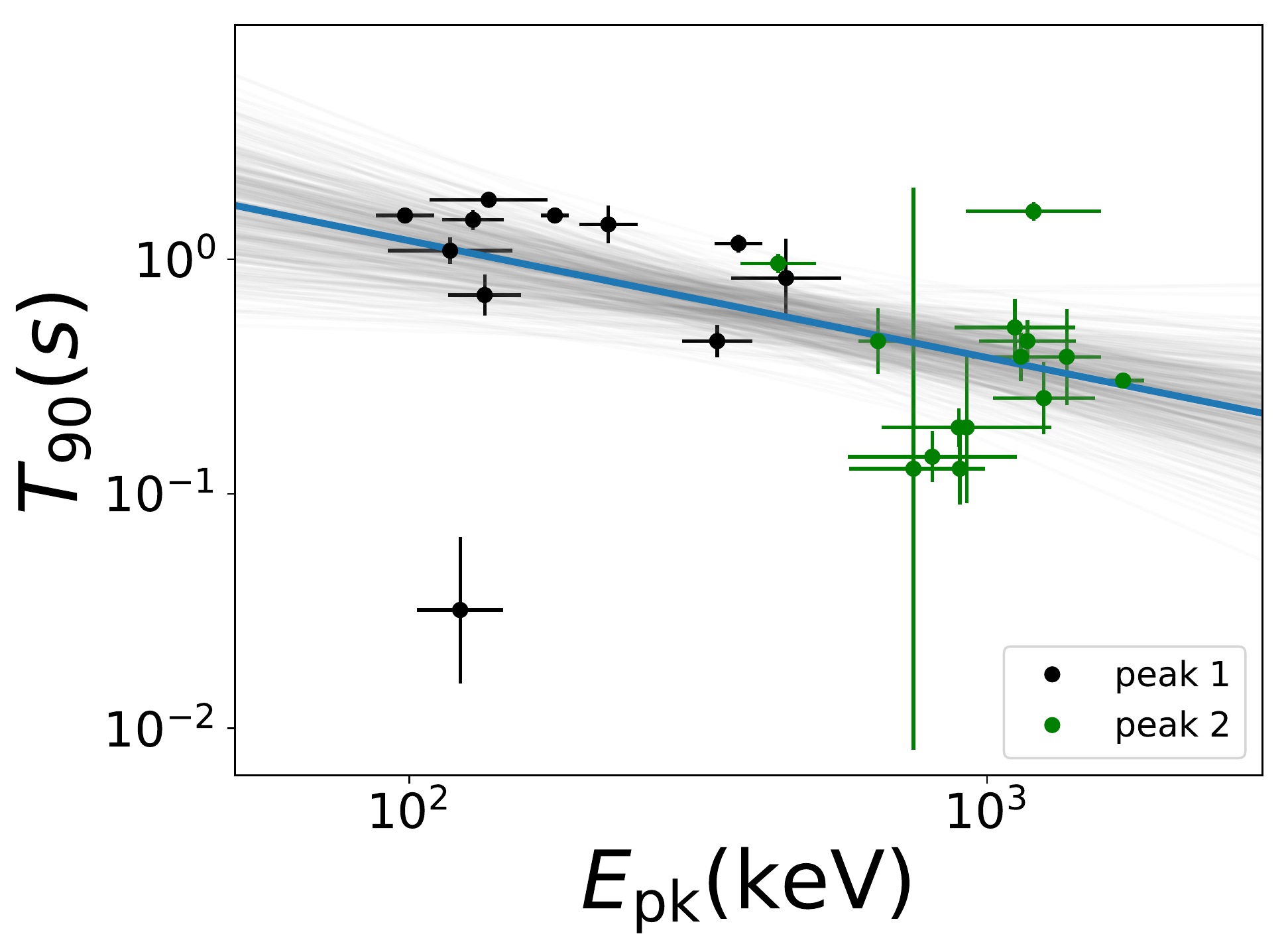}}
\subfigure{\includegraphics[align=t,width=0.45\textwidth]{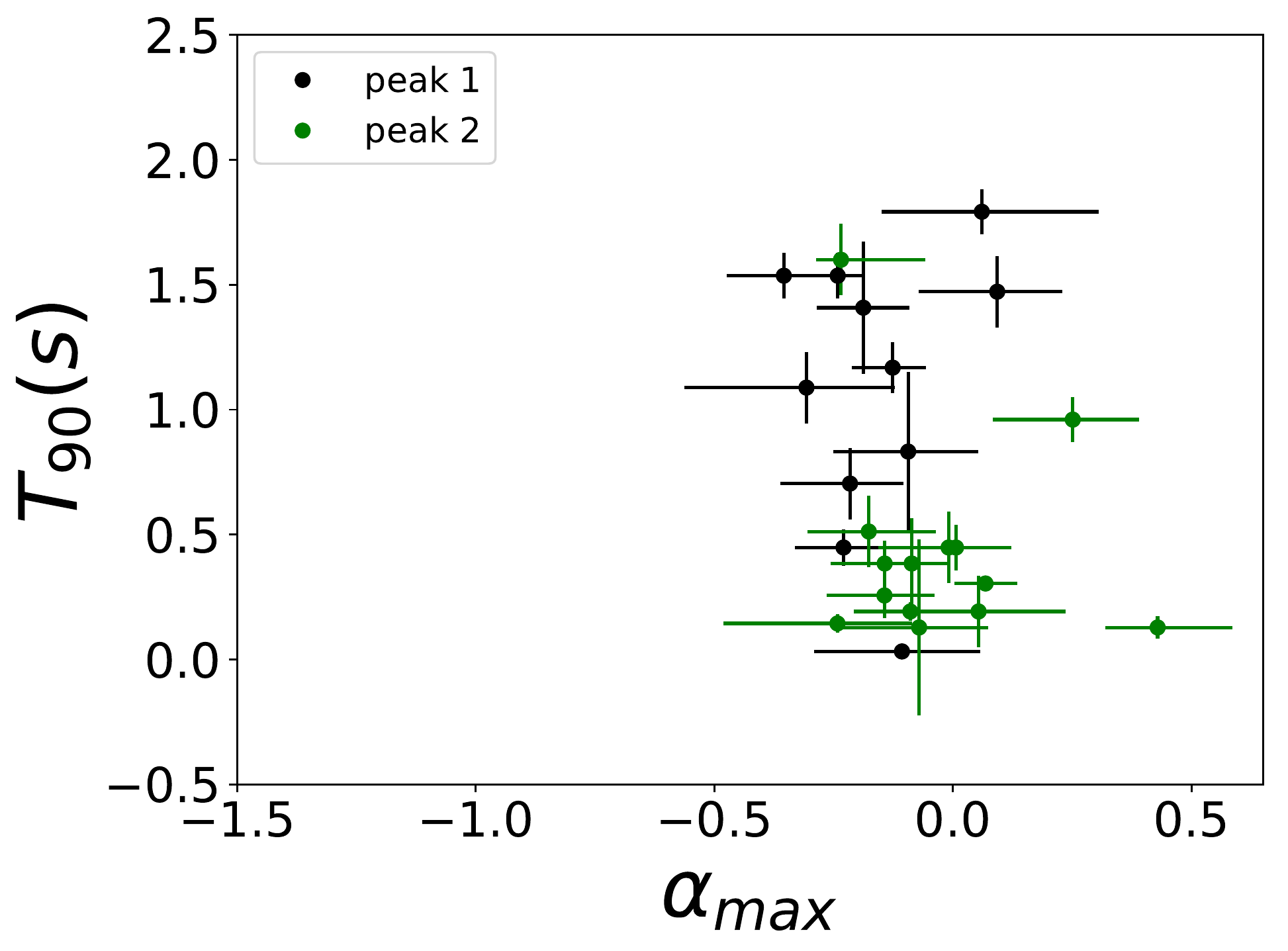}}
\caption{Parameter relations for the 25 pulses in our sample lying above the NDP line. The black and green data points correspond to the two peaks in the $\eta$ distribution defined in Figure \ref{fig:LFdistribution}.
Upper left panel: relation of $\eta$ versus $E_{\rm pk}$, which has the Spearman's rank correlation  coefficient $r  =  0.92$, corresponding to a chance probability of $p\ll0.00001$. Upper right panel: relation of $\eta$ versus $T_{90}$  ($r = -0.53$, $p = 0.01$). Bottom left panel: relation $T_{90}$ versus $E_{\rm pk}$ ($r = -0.43$, $p = 0.03$).  In this case, we fit for the correlation with a power law function $T_{\rm 90} \propto E_{\rm pk}^{-s}$ using Bayesian inference. The light blue line shows the mean of the posterior distribution (with a slope of $s=0.50$ and a corresponding standard deviation of $0.19$) and the grey lines are 1000 randomly selected samples from the Markov Chain Monte Carlo sampling, which shows the degree of spread in the posterior distribution of the slope.
Bottom right panel: relation $T_{90}$ versus $\alpha_{\rm max}$ ($r = -0.28$, $p = 0.18$).
\label{fig:LFdistribution1}}
\end{figure*}

For the two groups from the bi-modal $\eta$ distribution (Figure \ref{fig:LFdistribution}), a strong positive correlation between $\eta$ and $E_{\rm pk}$ is found (Figure \ref{fig:LFdistribution1}, top left panel). This can be explained as due to the computational dependence, as  $\eta \sim E_{\rm pk}^{1/2}F^{1/8}$ where $F$ is the flux in each of the analyzed timebins. The formula is adopted from \citet{Peer2007}, by identifying the peak energy with the black body temperature. The full dependence is $\eta \sim E_{\rm pk}^{1/2}F^{1/8}(1+z)^{1/2}d_L^{1/4}$. Therefore, it is important to note that the computation is not very sensitive to the uncertainty in the distance.

However, unexpectedly, we also find anti-correlations between $\eta$ and $T_{90}$ and between $T_{90}$ and $E_{\rm pk}$ (Figure \ref{fig:LFdistribution1}). The latter correlation is between observed quantities and thereby independent of any model for deriving their values, unlike~$\eta$. We therefore fit this correlation with a power law function $T_{\rm 90} \propto E_{\rm pk}^{-s}$ using Bayesian inference and account for the measurement errors in both parameters. We make use of Markov Chain Monte Carlo (MCMC) algorithms to explore the posterior distribution of the fit \citep[see, e.g.][]{Kelly2007}. This is shown by the grey lines (Figure \ref{fig:LFdistribution1}, bottom left panel) which are 1000 randomly selected samples from the MCMC sampling and shows the degree of spread in the posterior distribution of the slope. The blue line shows the mean of the posterior distribution and has a slope of $s = 0.50$. The corresponding standard deviation $0.19$.

\begin{figure*}
\centering
\subfigure{\includegraphics[align=t,width=\linewidth]{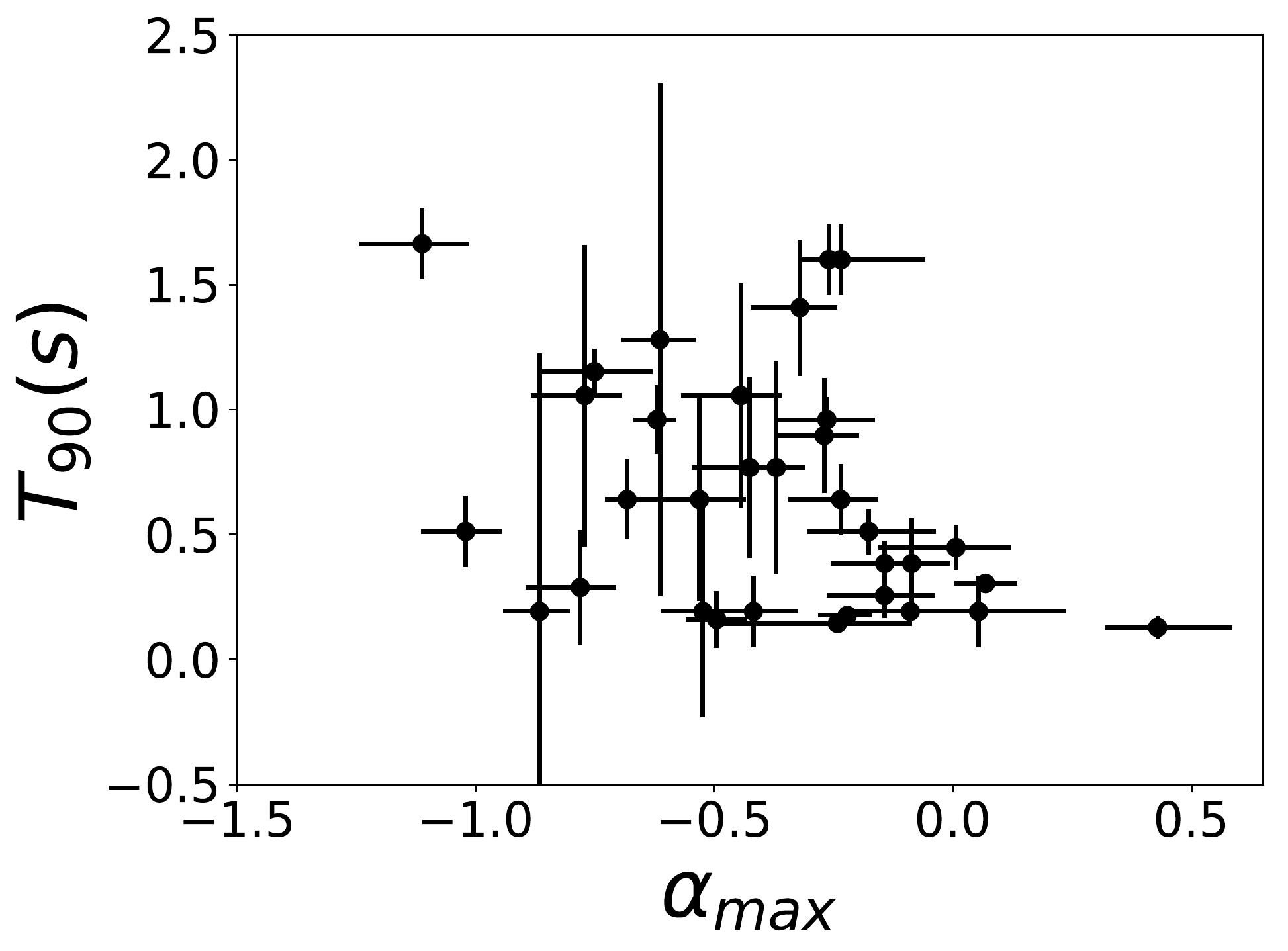}}
\caption{$T_{90}$ and $\alpha_{\rm max}$ relation for spectra with $E_{\rm pk} > 800$ keV in Figure \ref{fig:Epk_vs_alpha_max}. The Spearman's rank correlation  coefficient is $r = -0.37$, the chance probability is $p = 0.03$. 
\label{fig:T90_alpha_max_800keV}}
\end{figure*}

We do not observe any correlation between $T_{90}$ and the spectral slope $\alpha_{\rm max}$ for bursts above the NDP line (Figure \ref{fig:LFdistribution1}, bottom right panel). Similarly, no such correlation is found when considering the entire sample of 70 pulses. However, when we consider all the sources having  $E_{\rm pk} > 800$~keV (presented in Figure \ref{fig:Epk_vs_alpha_max}) we do observe an anti-correlation between $T_{90}$ and $\alpha_{\rm max}$ which is presented in Figure \ref{fig:T90_alpha_max_800keV}. This anti-correlation implies that sources with harder spectra have shorter $T_{90}$. 
This suggests that there might always exist a thermal emission at short times, which is accompanied by other emission processes such as synchrotron at later times. If this interpretation is correct, the lack of thermal emission in a given GRB might be explained by the lack of observed or studied spectra at sufficiently short time. 

\subsection{Hardness ratio}
\label{subsec:HR}
The anti-correlation we found between $E_{\rm pk}$ and $T_{90}$ of short GRBs with high $E_{\rm pk}$, motivated us to study a possible correlation between the hardness ratio (HR) and $T_{90}$. Following \citet{Kouveliotou1993}, we calculate the HR using the two typical energy bands, $100-300$ keV and $50-100$ keV. To integrate the spectra, we use the CPL fit parameters, $\alpha_{\rm max}$ and $E_{\rm c}$, for the 70 spectra in our sample. For comparison, we also calculated the HR for the spectra  with the maximal value of $\alpha$ in each of the 38 pulses from 37 long GRBs in the catalog by \citet{Yu2019}. These pulses were selected from single pulsed, long bursts that have at least 5 timebins in which the statistical significance is $S\geq20$.

\begin{figure*}
\centering
\subfigure{\includegraphics[align=t,width=0.45\linewidth]{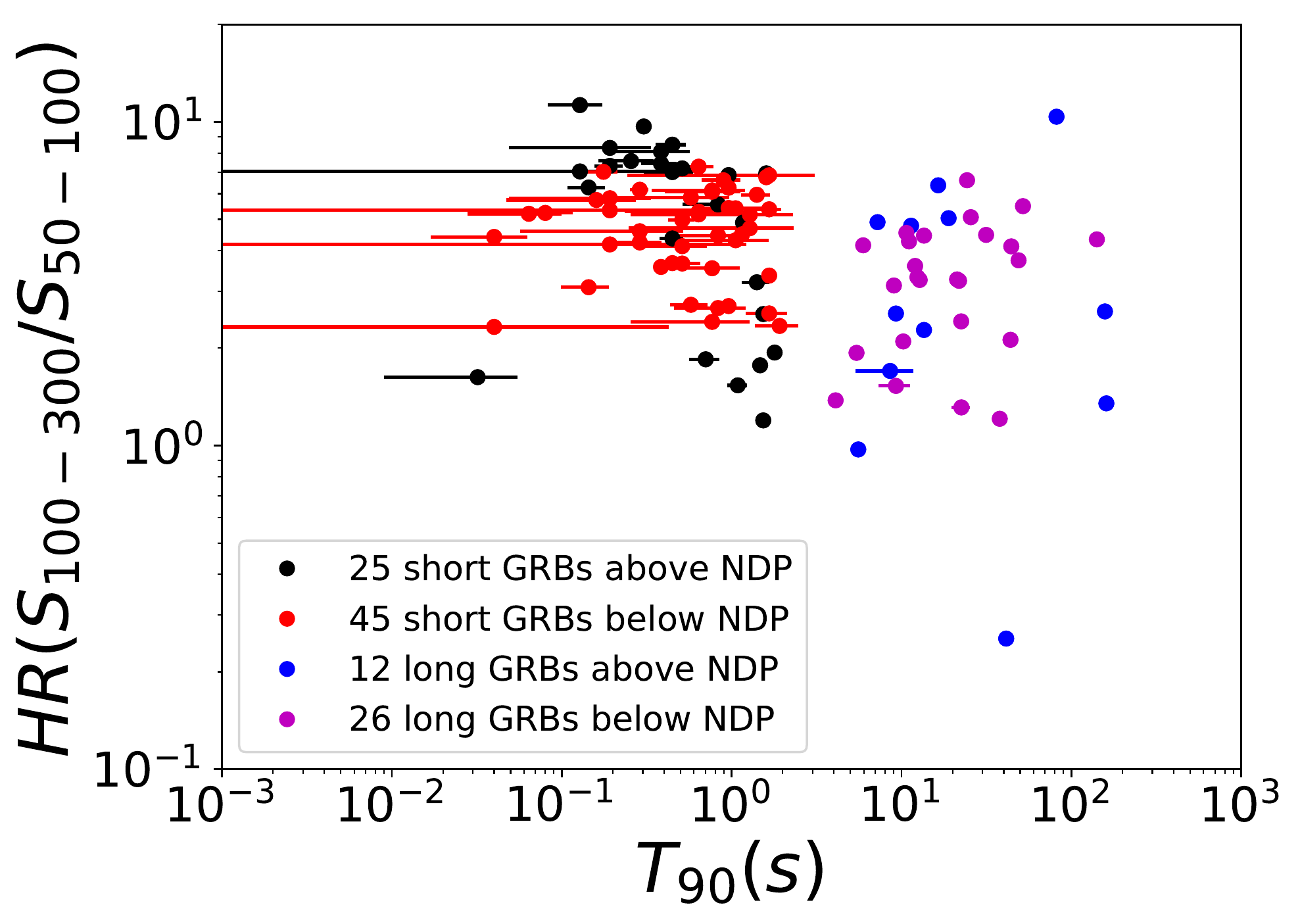}}
\subfigure{\includegraphics[align=t,width=0.45\linewidth]{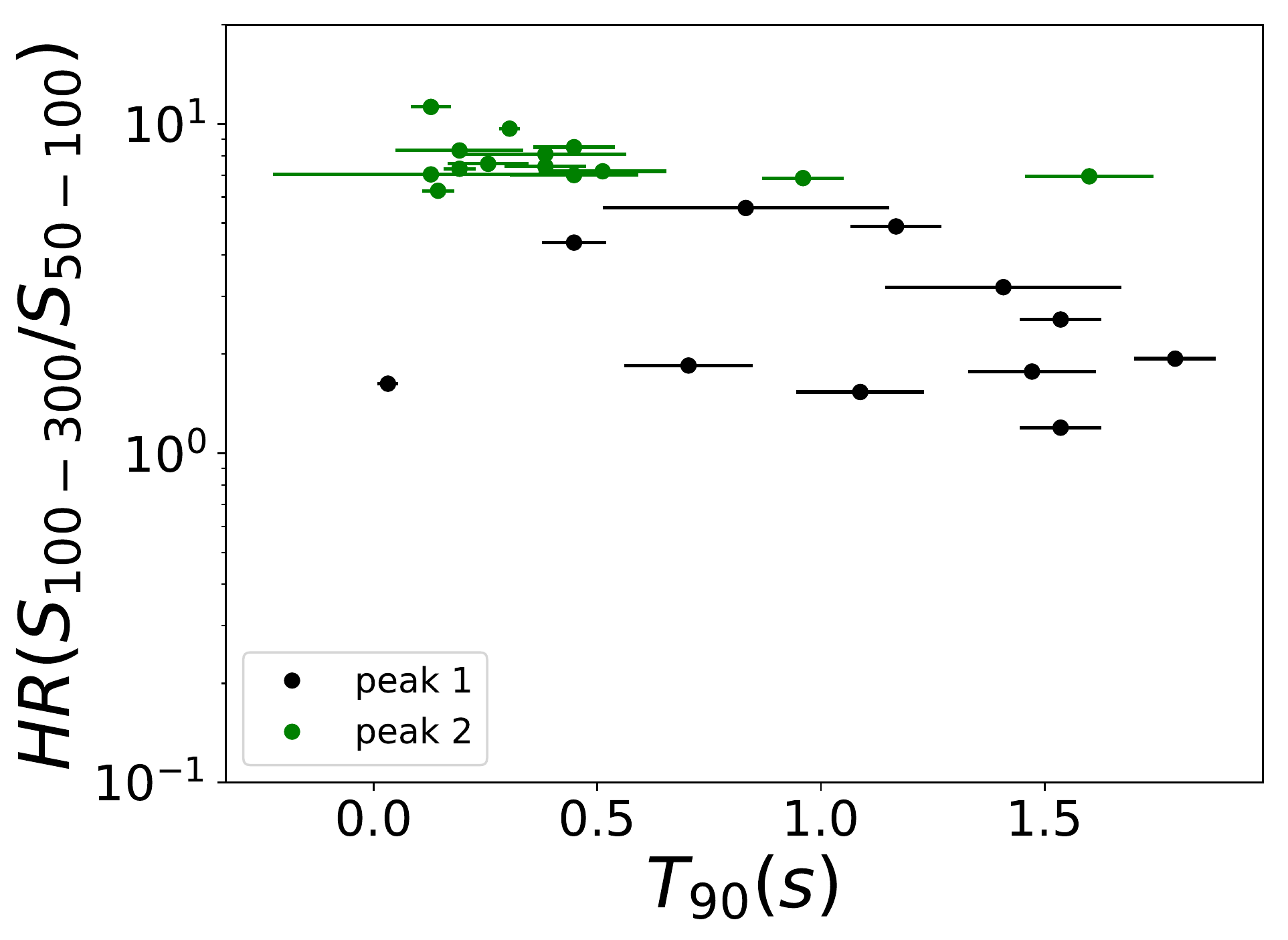}}
\caption{Hardness ratio (HR) versus pulse (burst) duration ($T_{90})$.
Left panel: 70 pulses from 68 short GRBs (red and black) and 38 pulses from 37 long GRBs (blue and purple). Right panel: 25 pulses from 24 short GRBs, all lying above the NDP line (the black and green data points correspond to the two peaks in the $\eta$ distribution defined in Figure \ref{fig:LFdistribution}). The Spearman's rank correlation coefficient is $r = -0.56$, the chance probability is $p = 0.01$.
\label{fig:HR_T90}}
\end{figure*}

The HR - $T_{90}$ relation is shown in Figure \ref{fig:HR_T90} (left panel), for both short and long GRBs. Short bursts with $\alpha_{\rm max}$ above the NDP model prediction are displayed in black while those below the prediction are in red. For the long bursts, those colors are blue and purple, respectively. While the $T_{90}$ selection criteria enables to clearly discriminate the long and short GRB population, we do not observe any additional correlation in this plot. 

The HR - $T_{90}$ relation for the 25 pulses from 24 short GRBs above the NDP line
is shown in the right panel in Figure \ref{fig:HR_T90}. The color coding (peak 1: black color and peak 2: green color) is the same as that used for the bi-modal $\eta$ distribution in Figure \ref{fig:LFdistribution}. Now, a clear separation is observed: GRBs with lower peak energy, have low Lorentz factor, lower hardness ratio, and longer $T_{90}$. 

Indeed, all the parameters of these GRBs in the first peak of the bi-modal $\eta$ distribution (in Figure \ref{fig:LFdistribution}, black color) seem to form a continuous distribution of the parameters of the population of long GRBs. This is in contrast to GRBs in the second peak of the bi-modal $\eta$ distribution (in Figure \ref{fig:LFdistribution}, green color), that have a higher HR than both the GRBs below the NDP line (45 pulses) as well as long GRBs (38 pulses). This result implies that the duration $T_{90}$ as a single criterion does not make a good separation between the two populations; rather as we show, the short GRBs may be composed of two separate populations - one which forms a continuation of the long GRB population, and another, separate population.

\subsection{Spectral parameter correlations for the two groups}
\label{subsec:parameter_correlations}
\begin{figure*}
\centering
\subfigure{\includegraphics[align=t,width=0.45\linewidth]{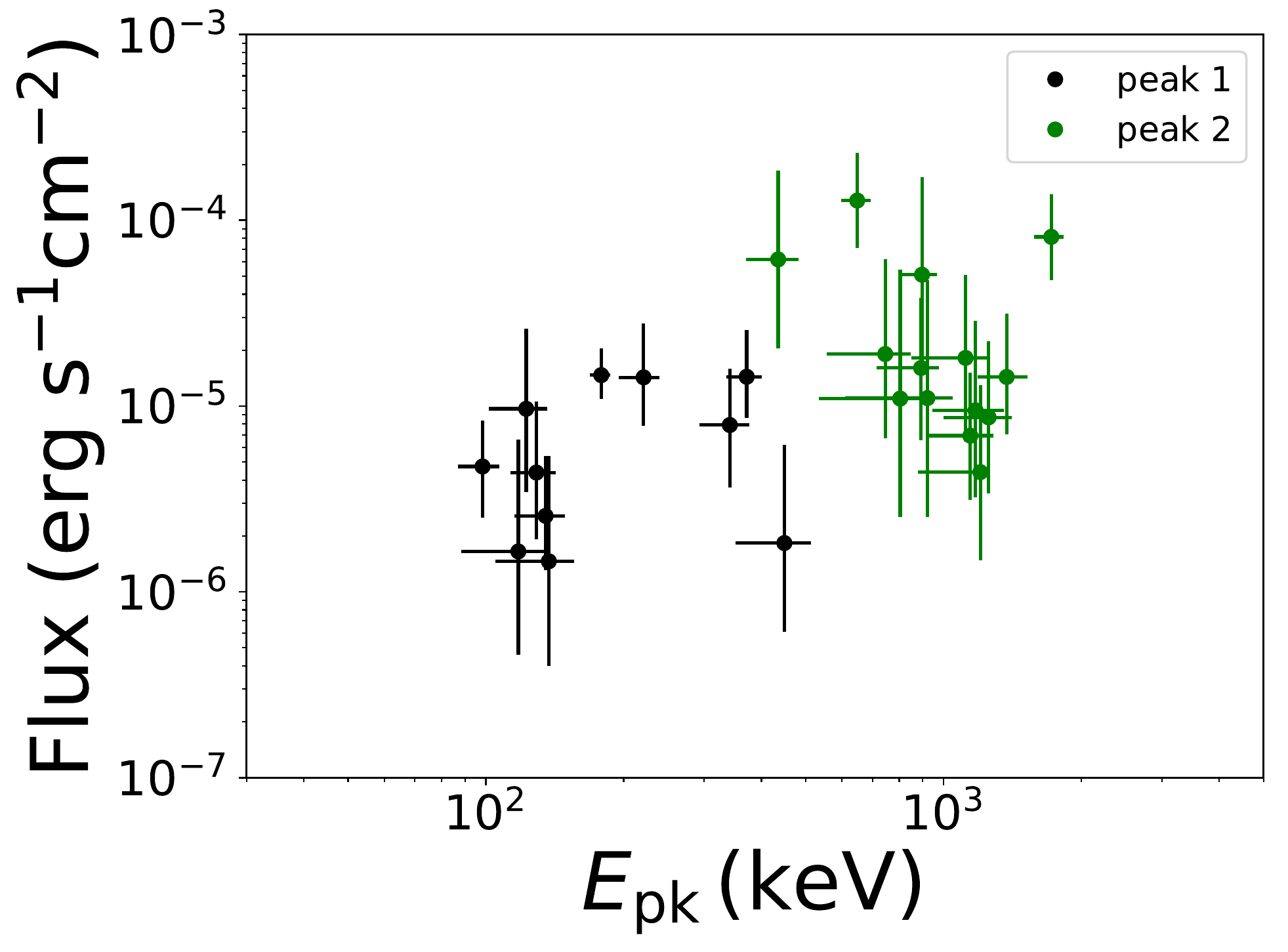}}
\subfigure{\includegraphics[align=t,width=0.45\linewidth]{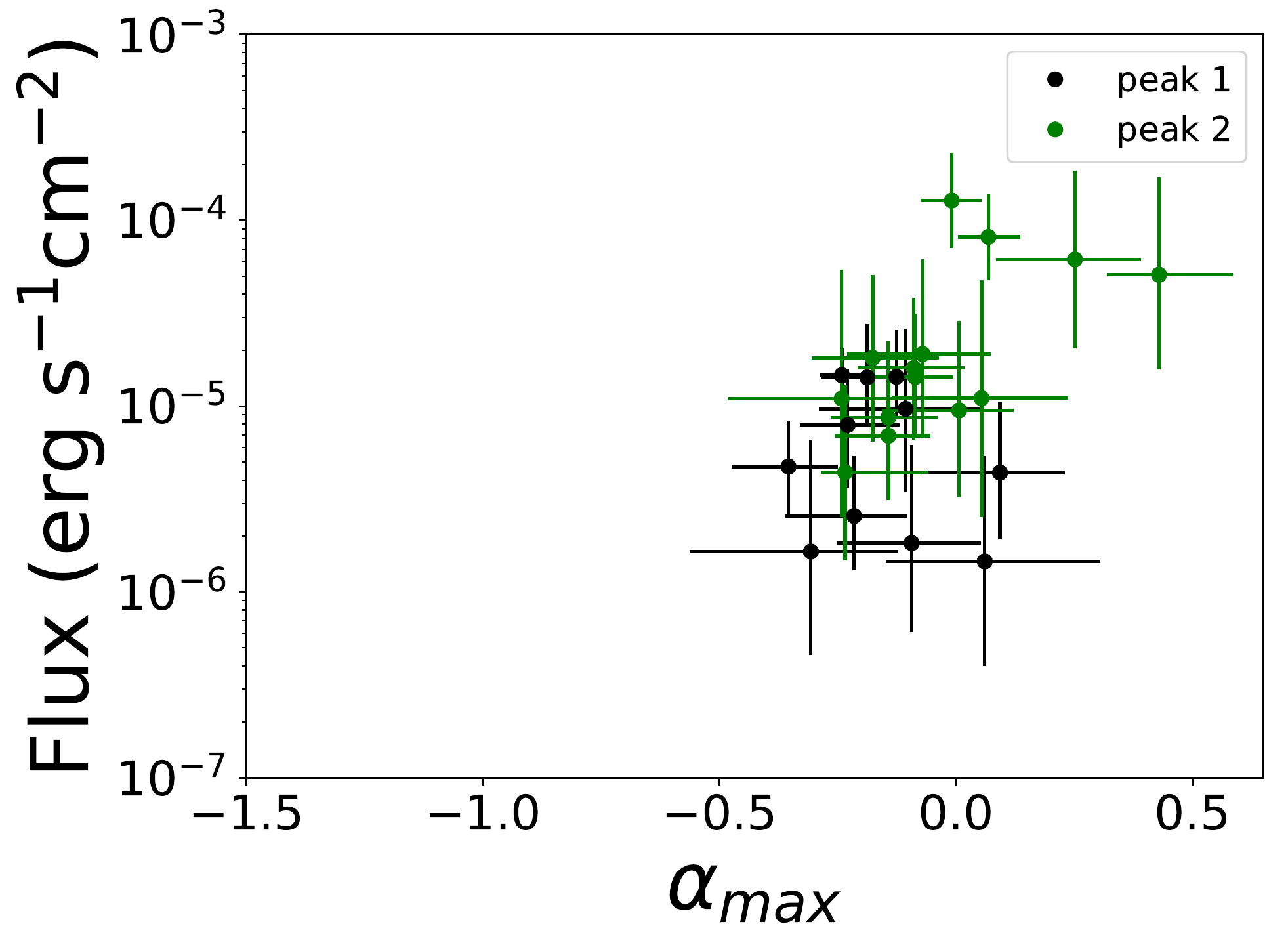}}
\caption{Flux dependencies for spectra above the NDP line (the black and green data points correspond to the two peaks in the $\eta$ distribution defined in Figure \ref{fig:LFdistribution}). Left panel: Flux versus  $E_{\rm pk}$ ($r = 0.40$ and $p = 0.05$). Right panel:  Flux versus $\alpha_{\rm max}$ ($r = 0.39$ and $p = 0.06$). 
\label{fig:flux_Epk_alpha}}
\end{figure*}

We find a weak positive correlations between the $F$ and $E_{\rm pk}$ (in Figure \ref{fig:flux_Epk_alpha}, left panel) and between the $F$ and $\alpha_{\rm max}$ (in Figure \ref{fig:flux_Epk_alpha}, right panel) for bursts  above the NDP line (in the two groups seen in the bi-modal $\eta$ distribution). However, no clear correlation is seen between these parameters for the bursts below the NDP line. This by itself is an interesting result. In the literature, several publications claim that there is a strong correlation between the luminosity, $L_{\rm peak}$ or the isotropic energy, $E_{\rm iso}$ and peak energy, $E_{\rm pk}$ \citep[e.g.,][]{Yonetoku2004, Amati2006, Ghirlanda2009}. These claims are based on a large sample of long GRBs. In contrast, here we do not find any such correlation when considering the entire sample of short GRB pulses, but we do find a correlation when we consider only those short GRBs whose spectral slope are above the NDP line.  

\section{Discussion}
\label{sec:Discussions}

\subsection{On the choice of the fitted model}

In this work, we fitted the \textit{Fermi}/GBM data using the phenomenological cut-off power law (CPL) model (this is also known as the "Comptonized" model). Several empirical models are commonly used in the literature for spectral analysis of GRBs. In addition to the CPL model \citep{Kaneko2006}, these include the "Band" model \citep{Band1993} as well as the smoothly broken power-law model \citep{Ryde1999}. \citet{Yu2016} showed that the CPL is the preferred model for the majority ($70\%$) of bursts, according to the Castor C-Statistic (CSTAT)\footnote{CSTAT is a modified version of the original Cash statistic \citep{Cash1979} in the case of Poisson data with Poisson background. Unlike the Cash statistic, it is used to determine an approximate goodness of fit measure to a given value of the CSTAT statistic.}. In addition, a consistent result was found by \citet{Yu2019} based on the Deviance Information Criterion (DIC) in Bayesian statistics \citep{Spiegelhalter2002}. This means that the results of the CPL model for these bursts have lower DIC and higher effective number of parameters, $p_{\rm DIC}>0$ \citep{Gelman2014} than those of the "Band" model. Additionally, the resulting parameters for the CPL fits are constrained within the prior ranges more often than those obtained from the BAND function fits; see also \citet{Burgess2019a, Burgess2019b}.  

It was recently argued by \citet{Burgess2019b} that a direct fit of the data to a synchrotron model enables to overcome the "line of death" criteria in many GRBs. However, this idea suffers several severe drawbacks. First, the bursts selected in that work are limited to bursts in the Yu et al. (2016) catalog, with the additional constraints of being single pulse GRBs and having known redshifts. This selection is different from the short pulses considered here. Second, the values of the parameters found in their fits require unacceptable high ratio of explosion energy to ambient mass density, of more than 7 orders of magnitude $(E/10^{53}\rm erg)/(n_{\rm ism}/1\rm cm^{-3})\gtrsim4\times10^7$ than the highest observed so far. In order to overcome this problem, \citet{Burgess2019b} suggested an additional acceleration of particles within the relativistically expanding jet ("jet within a jet"); however, no such mechanism that can lead to relativistic expansion within an already relativistically expanding jet is known. We thus find that this model is still incomplete, and an interpretation of an empirical fit still provides better insight. Another suggestion was given by \citet{Ghisellini2019} based on the low energy break in the prompt spectrum of GRBs. They argued that the emission process is still synchrotron radiation but produced by protons and cannot be completely cool. 


\subsection{Correlation between temporal and spectral structures} 

When we consider the entire sample of 70 pulses analyzed in the short GRB population, we do not observe a correlation between the burst duration $T_{90}$ and the peak energy, $E_{\rm pk}$. Similarly, for the long GRBs that we considered (both below and above the NDP line) no clear $T_{90}-E_{pk}$ correlation was detected. 
However, when we consider only those short bursts that have a hard value of the spectral index, $\alpha$, such that they are above the NDP line (Figures \ref{fig:Epk_vs_alpha_max}, \ref{fig:LFdistribution}) we do find an inverse correlation between $T_{90}$ and peak energy, $T_{90} \propto E_{\rm pk}^{-s}$, where slope of the power law function is $s = 0.50$ and the corresponding standard deviation is 0.19.

A quantitative relationship between the temporal and spectral structure in gamma ray bursts has been considered by several authors in the past. However, these works treated only the long GRBs. \citet{Richardson1996, Bissaldi2011} found a negative correlation between $T_{90}$ and peak energy, $T_{90} \propto E_{\rm pk}^{-s}$ where $s \simeq  0.4$. Their samples contained only bright, long GRB population. When considering the entire set of long GRB population,  \citet{Qin2013} report a similar correlation,  but with weaker dependence, $s =  0.2$. 

Here we report, for the first time, such a correlation in the sample of short GRBs. This could not have been done in the past, due to the small sample, or lack of suitable method to study the spectra.  The similarity between the correlation found here for short GRBs above the NDP line and for the bright long GRBs [which tend to have harder spectral index, $\alpha$; see \citet{Bissaldi2011}], as well as the fact that we do not detect any correlation for bursts below the NDP line, suggest a possible correlation between the emission mechanism and the burst duration. Bursts above the NDP line are consistent with originating from the photosphere, hence the photons directly probe the inner engine. While bursts whose spectra are below the NDP line may have additional radiative mechanisms, such as synchrotron emission, which originates from the outer regions of the outflow (outside the photosphere) and as such do not necessarily follow directly the duration of the inner engine. If this interpretation is correct, it points to a possible correlation between the duration of the inner engine and the temperature - or total energy, of the released photons. This further points to the importance of spectral analysis in analyzing possible correlations in the GRB population.  

A second correlation we find is between $T_{90}$ and $\alpha_{\rm max}$ when we consider a cut at higher peak energy ($E_{\rm pk} >800 \rm~ keV$) (see Figure \ref{fig:T90_alpha_max_800keV}). This (anti-) correlation further suggests a dual emission mechanism: short duration GRBs might be dominated by a thermal component, while an additional emission process may both lead to shallower spectra and be characterized by a longer duration.  
 
The results we find therefore strongly support the idea that the spectra of both short and long GRBs contain (at least) two separate components: a photospheric emission component that correlates directly with the inner engine activity, and a second component, possibly having a synchrotron origin, that is longer in nature, and less steep. 
 

\section{Summary and Conclusion}
\label{sec:Summary_Conclusion}

In this work, we have selected a sample of 70 pulses from 68 short GRBs with $T_{90}<2$ s detected by Fermi/GBM. These GRBs have at least one timebin with statistical significance $S\ge15$. The timebins were selected using the Bayesian block method that ensures that the intensity does not vary strongly during an individual timebin. A total of 153 time-resolved spectra were obtained and fitted with the empirical CPL spectral model, using a Bayesian statistical approach.

We investigate the distribution of the maximal (hardest) value of the spectral index $\alpha$ in each of the pulses, denoted $\alpha_{\rm max}$. Assuming that a single emission mechanism dominates throughout each pulse, the maximal value of the spectral index, $\alpha_{\rm max}$ provides a useful information on this emission mechanism. We find  that 70$\%$ (within a $1\sigma$ error) of short GRBs have at least one interval in which the value of $\alpha$ is beyond the value allowed by the "synchrotron line of death" (see Figure \ref{fig:alpha_max_histogram}). These values of $\alpha_{\rm max}$ are typically obtained when the flux is close to its peak (see Figure \ref{fig:example_LC}). Therefore, the emission mechanisms in these pulses are inconsistent with being dominated by synchrotron emission. 

When considering the intervals for which $\alpha = \alpha_{\max}$, we find that $36\%$ (within a $1\sigma$ error) of the spectra are consistent with having a non-dissipative photospheric origin, namely are above the NDP line \citep{Acuner2019}. This is presented in Figure \ref{fig:Epk_vs_alpha_max}. 
These numbers are slightly higher than that of long bursts. Indeed, short bursts have been found earlier to be harder than long bursts \citep{Kouveliotou1993, Tavani1998}. These results also prove the importance of using time resolved spectral analysis to access physical information of GRBs. 

For the bursts compatible with a non-dissipative photospheric origin, we calculate the coasting Lorentz factor, $\eta$, and find a bi-modal distribution in the values of $\eta$ (see Figure \ref{fig:LFdistribution}), peaking around $\sim 300$ and $\sim 1000$. The first peak ($\eta_{\rm pk,1} \sim 300$) is compatible with the average Lorentz factor $\eta$ found in long GRB population \citep{Racusin2011} while the second peak ($\eta_{\rm pk,2} \sim 1000$) is larger by a factor of $\gtrsim 3$. 

A clear separation between bursts that belong to these two distinct peaks in the  $\eta$ distribution is further observed in their duration ($T_{90}$), peak energies ($E_{\rm pk}$) and hardness ratio (see Figure \ref{fig:HR_T90}).
For these bursts, we further find a strong positive correlation between $\eta - E_{\rm pk}$ and a negative correlation between $T_{90} \propto E_{\rm pk}^{-s}$ with a power law index $s=0.50$ and a corresponding standard deviation 0.19 (see Figure \ref{fig:LFdistribution1}).

We also find an anti-correlation between $T_{90}$ and $\alpha_{\rm max}$ when we consider a cut at large peak energy ($E_{\rm pk} >800 \rm~ keV$), see Figure \ref{fig:T90_alpha_max_800keV}. This indicates that here in our sample most pulses are compatible with thermal emission at short times but with some contamination from other emission processes such as synchrotron at later times.

The bi-modal distribution we find in the values of the Lorentz factor, together with the differences in the harness ratio, provide a strong indication that what is currently classified as short GRBs, in fact is made of two separate populations. The first is an extension of the long GRB population to shorter duration, and the second is a truly separated population. A striking result is the difference in the Lorentz factors, by an average factor of $\gtrsim 3$, with this separate population having Lorentz factor of $\sim 1000$, and in some cases higher. This implies that, on the average, the outflows of the separate population contain much less ejected material than that of long GRBs, which provides a further clue to the true nature of short GRB progenitors.  

Our results provide a direct indication that the GRB duration by itself is not sufficient to classify the nature of a GRB: $T_{90}\lesssim2 s$ or $T_{90}\gtrsim2 s$ by itself is not enough to separate short and long GRBs. Rather, one needs to consider additional information, which includes a spectral information, such as the hardest value of $\alpha$ in each pulse/burst, and the corresponding Lorentz factor, $\eta$.   
Indeed, classification of GRBs is long discussed in the literature as a way of discriminating GRB progenitors \citep[][and references there in]{Kouveliotou1993, Tarnopolski2015}. Here we show that the maximal (hardest) value of $\alpha$ in each pulse/burst can be used as an additional method for the classification of bursts, especially for the classification of short ones.

These surprising results lead us to conclude the following: (1) a thermal (photospheric) emission is ubiquitous among short GRBs, with $\sim 1/3$ being consistent with having a pure thermal origin, and another large fraction may also have a thermal origin, which is distorted by sub-photospheric energy dissipation. However, this component is often accompanied by an additional emission mechanism (likely, synchrotron) which makes it hard to separate and clearly identify the dominant mechanism. (2) At early (short) times, the thermal component often dominates, but at longer times it is accompanied by a second mechanism, which makes it sub-dominant. (3) Only for those bursts in which the thermal component dominates, we find a correlation between pulse (burst) duration $T_{90}$, and the peak energy, $E_{\rm pk}$ which corresponds to the temperature: higher peak energy correspond to shorter burst duration. Since no corresponding correlation is found in the flux, this implies that a similar amount of energy is released in short time, which leads to higher temperature. This result may therefore provide a very strong hint towards a better understanding of the progenitor models and explosion mechanisms in short GRBs. (4) When considering only those bursts with high peak energy, $E_{\rm pk} >800 \rm~ keV$, we further find a correlation between the burst duration $T_{90}$ and the hardest spectral slope $\alpha_{\rm max}$. This further supports the idea of a dual emission mechanisms: thermal and non-thermal (synchrotron). 



\acknowledgments
We wish to thank Dr. Zeynep Acuner for enlightening discussions on the manuscript. We also wish to thank Drs. Jochen Greiner and Hoi-Fung Yu for comments. This research made use of the High Energy Astrophysics Science Archive Research Center Online Service HEASARC at the NASA/ Goddard Space Flight Center.  We acknowledge support from the Swedish National Space Agency (196/16), the Swedish Research Council (Vetenskapsr\aa det, 2018-03513), and the Swedish Foundation for international Cooperation in Research and Higher Education (STINT, IB2019-8160). F.R. is supported by the G\"oran Gustafsson Foundation for Research in Natural Sciences and Medicine. A.P. is partially supported by the European Research Council via ERC consolidating grant 773062 (acronym O.M.J.).

%
\vspace{5mm}
\facilities{{\it Fermi}/GBM}
\software{{\tt 3ML} \citep{Vianello2015}}





\appendix
\section{Selection of statistical significance}
\label{app:significance_level}
We used the criterion that the significance $S$ should be larger than $15$ for each timebin that was analysed and interpreted.   This criterion was found adequate to ensure that the spectral slopes and peak energies are determined with sufficient accuracy.
In order to find the appropriate level of significance, we generated a large number of synthetic spectra, with different peak energies. 
The properties of the simulated observations, such as the detector response and viewing angle, were based on the observations of GRB090820 (which had $\alpha$ = -0.5) and the background spectrum was assumed to be a power-law with index -1.5.
The normalization of these generating models was chosen  such that the significance for each spectrum attained the same value. 
This process was repeated for three cases $S = 10, 15$, and $20$ and the distributions of the fitted spectral parameters were compared. The results showed similar parameter distributions for the two cases with $S = 15$ and $20$, however, the spectra got softer at lower values of  $S$. This can be explained by the limited instrumental energy range which does not allow to properly capture the low energy spectral slope, when the data has a too low value of $S$. In fact, we found that, even at low values of $S$, a large spectral peak energy still allowed us to determine the low-energy index correctly. Since short GRBs have, on average, larger spectral peak energies compared to long GRBs \citep[e.g.,][]{Ghirlanda2011}, the criterion $S\ge15$ can be used instead of the criterion $S\ge20$, which was, e.g., used for long bursts in \citet{Yu2019}. Therefore, the limit of $S\ge15$ is necessary and enough to suppress instrumental effects and was therefore used to ensure well-constrained spectral fits in our sample of short GRBs.


\end{document}